\documentclass[11pt]{article}
\usepackage[a4paper,margin=2.5cm]{geometry}
\usepackage[T1]{fontenc}
\usepackage{graphicx}
\usepackage{amsmath}
\usepackage{booktabs}
\usepackage{multirow}
\usepackage{placeins}
\usepackage{siunitx}
\usepackage{gensymb}
\usepackage{authblk}
\usepackage[hidelinks]{hyperref}

\DeclareSIUnit{\clight}{\ensuremath{c}}

\title{Experimental Demonstration of Soft-Tissue Muon Scattering Tomography}

\author[1]{D.~Banerjee\thanks{Corresponding author: \texttt{dipanwita.banerjee@cern.ch}}}
\author[2]{B.~Banto~Oberhauser}
\author[2]{P.~Crivelli}
\author[3]{L.~Molina~Bueno}
\affil[1]{CERN, European Organization for Nuclear Research, 1211 Geneva 23, Switzerland}
\affil[2]{Institute for Particle Physics and Astrophysics, ETH Z\"urich, 8093 Z\"urich, Switzerland}
\affil[3]{Instituto de F\'{\i}sica Corpuscular (CSIC/Universitat de Val\`encia), 46980 Paterna, Valencia, Spain}

\date{\today}

\begin{document}
\maketitle

\begin{abstract}
Muons have fascinated physicists since their discovery in cosmic rays almost a century ago. Their large mass and relatively long lifetime allow relativistic muons to traverse large amounts of matter, a property that has been exploited to probe otherwise inaccessible structures, from volcanoes and archaeological monuments to shielded nuclear materials. Here, we experimentally demonstrate for the first time that muon scattering tomography can be extended to biological soft tissue. We image an anthropomorphic thorax phantom using a \qty{2}{\giga\electronvolt\per\clight} muon beam at the CERN Proton Synchrotron and reconstruct its three-dimensional scattering density from the measured deflections of individual muons. Despite the weak scattering produced by tissue-equivalent materials, the reconstructed density agrees with the expectation from simulation at the percent level. Our results extend muon imaging into a previously unexplored regime and open the possibility of using muons as a complementary probe for medical imaging.

\end{abstract}

\section{Introduction}

The discovery of the muon has an intriguing history. Muons were first observed in cosmic rays in the 1930s~\cite{neddermeyer1937,street1937} and were initially believed to be the meson predicted by Yukawa to mediate the nuclear force \cite{yukawa1935}. Their mass, about 200 times that of the electron, appeared to make them a natural candidate. It soon became clear, however, that they interacted far too weakly with nuclei to play this role~\cite{conversi1947}. The particle that had been discovered was instead a heavier relative of the electron, prompting the famous question attributed to Isidor Rabi, ``Who ordered that?''. Its weak interaction with matter explained why muons constitute the penetrating component of cosmic radiation that reaches sea level. The pion, which does participate in the strong interaction and is associated with Yukawa’s original idea, was identified only later in photographic emulsions exposed to cosmic rays at high altitude~\cite{lattes1947}.

Since then, the muon has played an important role across particle, nuclear, atomic and condensed-matter physics. As a second-generation lepton, its properties provide stringent tests of the Standard Model and sensitive probes for physics beyond it. Precision measurements of its lifetime ~\cite{webber2011mulan} and anomalous magnetic moment ~\cite{muong2final}, searches for charged-lepton-flavour violation ~\cite{calibbi2018clfv}, and the spectroscopy of muonic atoms ~\cite{schuhmann2025helion} and muonium~\cite{janka2022muonium,strasser2025museum} continue to exploit this apparently simple particle to test fundamental physics at increasingly high precision. High-energy muon beams also provide a direct probe of new weakly coupled particles and interactions~\cite{andreev2024na64mu}.
At the same time, implanted spin-polarised muons provide a sensitive local probe of matter through muon spin rotation, relaxation and resonance ($\mu$SR), with applications ranging from magnetism and superconductivity to semiconductors, battery materials and ionic diffusion~\cite{hillier2022musr}.

The penetrating nature of the muon, already evident from its role in cosmic radiation, is also one of its most remarkable practical properties. Owing to their relatively large mass, radiative energy losses are strongly suppressed compared with electrons, enabling relativistic muons to traverse large thicknesses of matter with comparatively little energy loss~\cite{PDG2024}. Nature itself provides a continuous source of such particles. Cosmic rays interacting in the atmosphere produce pions and kaons that decay into energetic muons, many of which reach the Earth's surface. The same penetrating power that played a central role in the early identification of the muon has therefore turned it into a unique probe of otherwise inaccessible structures~\cite{bonechi2020}.

This idea has given rise to muon imaging, or muography. In transmission muography, the attenuation of the cosmic-ray muon flux is used to infer the density distribution of an object. The technique has been applied to geological structures, volcanoes, glaciers, civil engineering and archaeology~\cite{bonechi2020,procureur2018}, and famously revealed a previously unknown large void inside the Great Pyramid of Giza~\cite{morishima2017}. A complementary approach, muon scattering tomography (MST), measures instead the angular deflections that individual muons undergo while traversing matter~\cite{borozdin2003,schultz2007}. Over the past two decades MST has developed into a powerful technique for imaging dense, high-$Z$ materials, with applications ranging from cargo inspection and nuclear safeguards to reactor monitoring and the search for shielded nuclear material. Several commercial systems based on these principles are already in operation~\cite{bonechi2020,procureur2018}.

The imaging principle underlying MST is multiple Coulomb scattering. As charged particles traverse matter they undergo repeated Coulomb interactions with atomic nuclei, resulting in a characteristic angular broadening approximately given by~\cite{highland1975,lynch1991}

\[
\theta_0 \propto \frac{\sqrt{L/X_0}}{p},
\]

where $L$ is the traversed thickness, $X_0$ is the radiation length of the material, and $p$ is the muon momentum. The scattering is therefore strongly dependent on the material traversed. This explains the historical success of MST for dense and high-$Z$ objects, where large scattering contrasts are available.

Biological tissue represents a very different regime. Soft tissues consist predominantly of low-$Z$ elements and differ only moderately in density, producing scattering signals orders of magnitude smaller than those typically exploited in conventional muon tomography. Extracting three-dimensional information in this regime requires precise tracking and sufficient statistics to resolve small differences in the scattering distribution. Whether muon scattering tomography can be extended from its traditional high-contrast applications to biological soft tissues has therefore remained an open question.

The possibility is nevertheless intriguing. GeV muons traverse the human body while depositing energy predominantly through minimum ionisation, and their trajectories can be measured individually before and after passage through the object. Muon imaging would not be expected to replace established techniques such as X-ray computed tomography (CT), but its intrinsically small energy deposition could make it particularly interesting for applications in which low radiation exposure or repeated imaging is advantageous. As discussed below, the muon fluence used in the fully sampled region of the present measurement corresponds to a local absorbed dose of only a few $\mu$Gy, illustrating the low-dose potential of the technique. Until recently, however, such a possibility remained largely academic. The natural cosmic-ray muon flux is far too low for practical medical imaging, while intense and well-controlled muon beams have traditionally been available only at large accelerator laboratories.

This situation may evolve with the development of new high-intensity and potentially more compact muon sources. Accelerator-based concepts~\cite{xu2025hiafmuon}, fixed-field alternating-gradient machines~\cite{machida2012emma} and high-intensity pion-production targets~\cite{knecht2021himb} are being explored to increase the available muon flux. Particularly interesting are laser-plasma accelerator concepts, in which GeV electron beams generated over centimetre-scale plasma stages can produce directional muon beams through pair production in a converter. Such a source has recently been demonstrated experimentally at the Berkeley Lab Laser Accelerator~\cite{lblmuons2025}. Although these approaches do not yet provide the flux and beam quality available at dedicated accelerator facilities, they motivate the question of whether muon imaging can be extended into regimes that have so far remained unexplored.

Studies of muon imaging for biological targets remain scarce. Morris et al. recently demonstrated cosmic-ray radiography of a human phantom using transmission, stopped-track and multiple-scattering techniques, obtaining two-dimensional radiographs from a 24-hour exposure. They proposed continuous, radiation-free monitoring of bulk anatomical parameters such as lung density in non-ambulatory patients, using only the natural muon background~\cite{morrisphantom2025}. Here we demonstrate, to our knowledge, the first experimental realisation of muon scattering tomography in the soft-tissue regime. Using a \SI{2}{GeV/\clight} positive-muon beam at the CERN Proton Synchrotron, we image an anthropomorphic thorax phantom from seven projection angles and reconstruct its three-dimensional scattering-density distribution from the measured deflections of individual muons. Despite the weak scattering produced by tissue-equivalent materials and the limited angular coverage and statistics of the beam test, the reconstructed soft-tissue density agrees quantitatively with the expectation from simulation. Our results demonstrate that the weak-scattering regime characteristic of biological tissue is experimentally accessible to muon tomography and provide a basis for exploring its potential for future medical imaging applications.

\section{Results}

\begin{figure}[htbp!]
    \centering
    \includegraphics[width=0.85\linewidth]{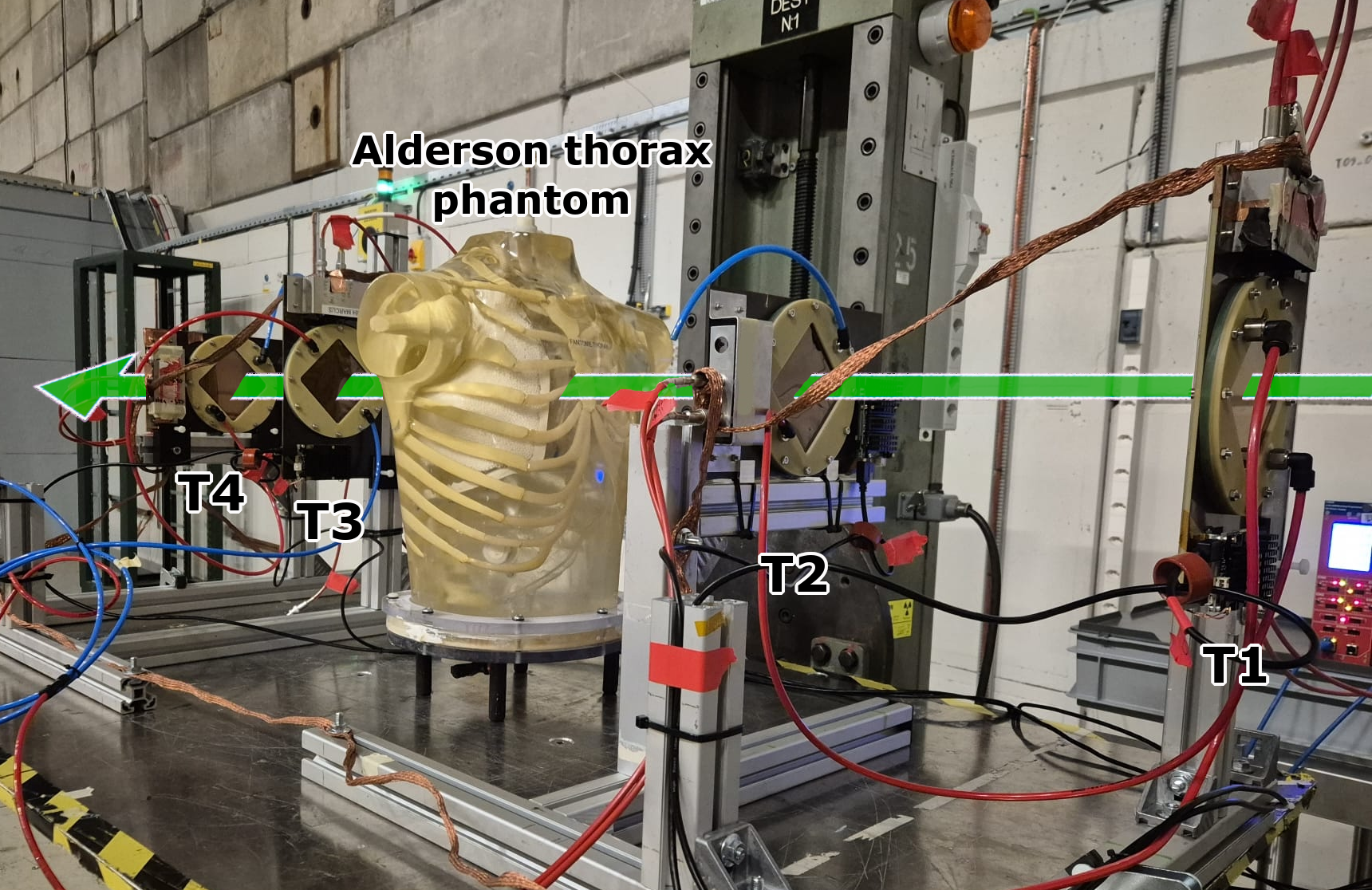}
    \caption{The Alderson thorax phantom installed on the rotation stage between the upstream (T1, T2) and downstream (T3, T4) Micromegas tracker pairs at the T9 beamline. The green arrow indicates the muon beam trajectory.}
    \label{fig:phantom_photo}
\end{figure}

We imaged an anthropomorphic Alderson thorax phantom~\cite{rsdtechsheet} which is a clinical dosimetry phantom built from tissue, lung, bone and blood-equivalent materials, with a \qty{2}{\giga\electronvolt\per\clight} positive-muon beam at the T9 beamline of the CERN Proton Synchrotron~\cite{bernhard2019eastarea}. Four Micromegas trackers~\cite{giomataris1996, Banerjee:2017mdu} with an active area of $8 \times 8$~\unit{\centi\metre\squared} each, two upstream and two downstream of the phantom as shown in Fig.~\ref{fig:phantom_photo}, recorded the incoming and outgoing trajectory of every muon, from which its projected scattering angle was obtained. The phantom was rotated through seven projection angles between \ang{0} and \ang{270}. The three-dimensional scattering-density map $\lambda$ is reconstructed in two stages (full details in Section~\ref{sec:methods}): a maximum-likelihood expectation--maximisation (MLEM)~\cite{schultz2007,sheppvardi1982} inversion of the measured scattering angles produces an unbiased but noisy volume, which is then cleaned by a three-dimensional U-Net denoiser~\cite{ronneberger2015,cicek2016} trained \emph{exclusively} on Monte Carlo simulations.

The training set is generated with BDSIM~\cite{nevay2020} (Geant4-based~\cite{geant4}), using a parametric model of the Alderson Heart \& Thorax phantom built from the manufacturer's specifications~\cite{rsdtechsheet}. Fifty phantom geometries were generated with per-seed randomisation: independent $\pm 10\%$ linear scaling of each organ, $\pm 10\%$ rib size, and random repositioning of the three defects within the myocardial wall. Each geometry carries an analytic ground-truth (GT) $\lambda$ map on its own grid.

Because the T9 muon beam is a narrow, structured spot (roughly \qtyproduct{60 x 40}{\milli\metre}) and each Micromegas covers a small acceptance, only a limited tube through the phantom is sampled at each projection angle. Three illumination scenarios were therefore tested. (i)~\textbf{Full-coverage illumination} is an idealised, uniform flat field that floods the entire phantom cross-section; it exists only in simulation and measures the best-case, intrinsic performance of the reconstruction. (ii)~\textbf{Acceptance-matched illumination} restricts the simulation to the \emph{measured} per-angle beam footprint and the real detector acceptance, so it reproduces exactly what the beam sees during the scan; here only the sub-volume crossed by the beam at \emph{all} seven angles, which we term the \emph{common core}, is fully sampled and quantitative. (iii)~The \textbf{real beam data} are the measured scan, to which the acceptance-matched, simulation-trained network is applied. All quantitative comparisons are made in the common core, defined next.

\subsection{Region of the phantom imaged by the beam}
\label{sec:imaged}

\begin{figure}[htbp!]
    \centering
    \includegraphics[width=\linewidth]{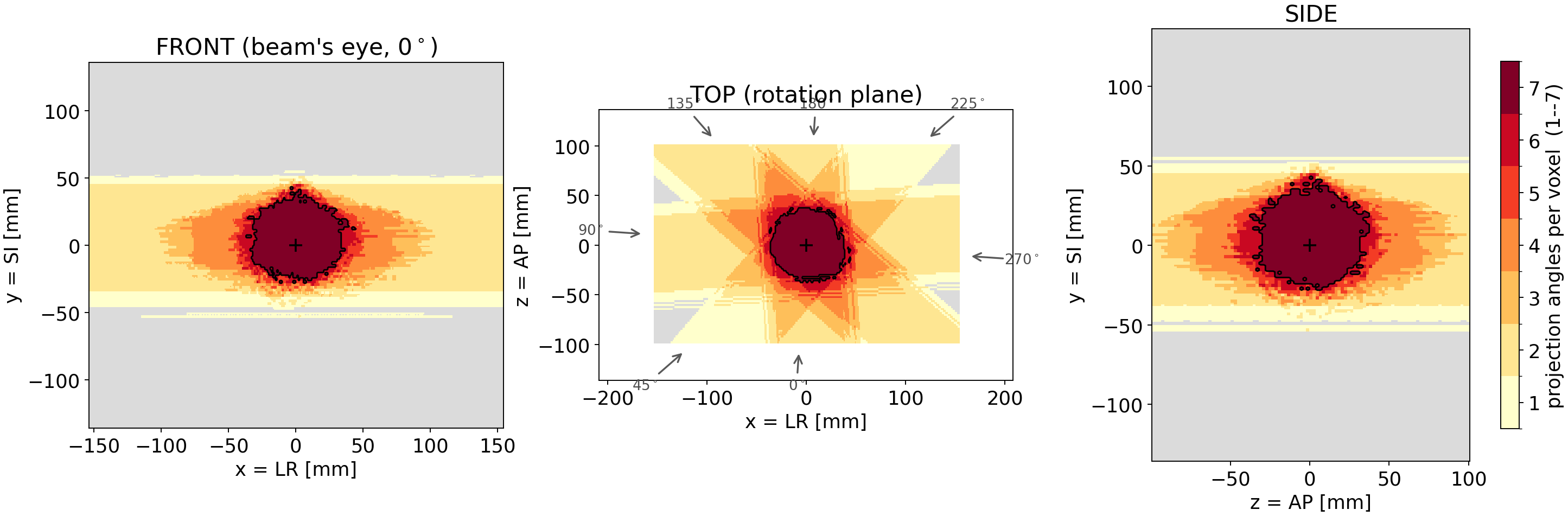}
    \caption{Imaged region of the phantom under the real-beam acceptance (front, top and side views): body (grey), voxels imaged at $\geq 1$ angle (light blue, $\approx 20\%$ of body) and the common core imaged at all seven angles (dark blue, $\approx 0.8\%$ of body), with the phantom centre marked. Grey arrows indicate the seven beam directions.}
    \label{fig:imaged}
\end{figure}

Figure~\ref{fig:imaged} quantifies which voxels are actually sampled under the real acceptance. The region of the phantom reached at one or more angles covers only $\approx 20\%$ of the body and forms a characteristic seven-arm star in the transverse plane. The \emph{common core} is the region of the phantom reached at \emph{all} seven angles and is the section where the tomographic problem is fully determined. It is the intersection of these tubes with the body: approximately 0.8\% of the body volume, a roughly cylindrical region about \qtyproduct{80 x 70 x 80}{\milli\metre} around the phantom centre. Anatomically this is the heart/mediastinum region, which in the Alderson phantom is homogeneous tissue-equivalent material containing no bone. All quantitative statements below refer to this common core; structures outside it (ribs, sternum, spine and lung boundaries) are only partially sampled and, while they appear at roughly their correct positions, their absolute $\lambda$ values are not fully constrained.

\subsection{Validation of the reconstruction on simulated phantoms}
\label{sec:performance}

\begin{figure}[htbp!]
    \centering
    \includegraphics[width=0.95\linewidth]{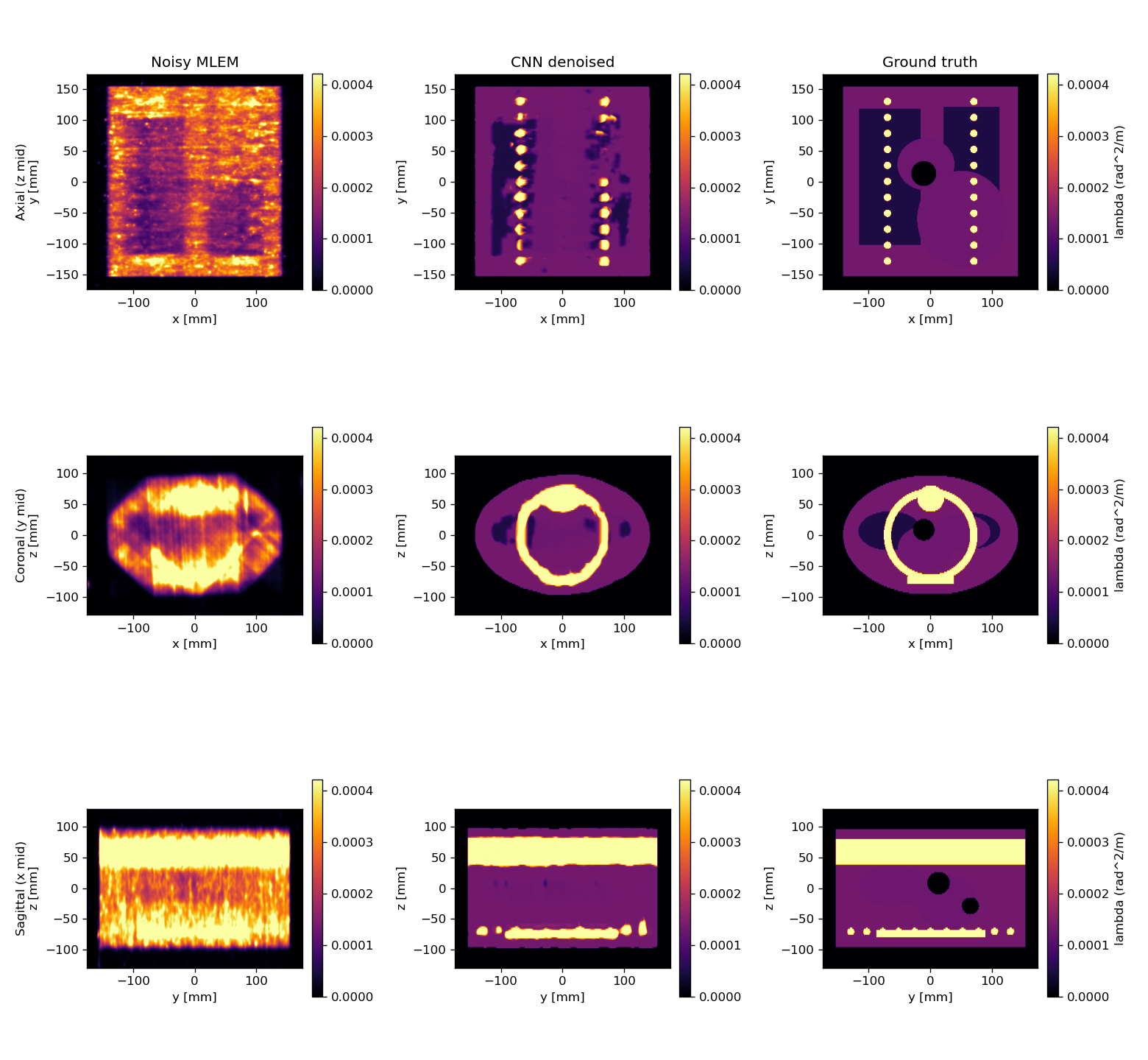}
    \caption{Independent evaluation phantom (seed 9000), full-coverage illumination: axial, coronal and sagittal sections of the noisy MLEM reconstruction (left), the CNN output (centre) and the ground truth (right). Body RMSE improves by a factor 3.0. Lungs, rib cage, sternum, spine and myocardium are recovered; the small air-filled myocardial defects are not, as expected at these statistics.}
    \label{fig:seed9000}
\end{figure}

We first establish the intrinsic capability of the reconstruction under full-coverage illumination, where the whole phantom is uniformly sampled. On five held-out simulated phantoms, unseen during training, the denoiser reduces the in-body voxel-wise root-mean-square error (RMSE) between reconstruction and ground truth from $\approx 1.5\times10^{-4}$ (MLEM alone) to $\approx 5.2\times10^{-5}$, a consistent factor of 2.8--3.1, and the anatomy --- lungs, rib cage, sternum, spine, myocardium and liver --- is recovered. As a stricter test of generalisation, a completely independent phantom (a random geometry never used in training, validation or testing) passes through the identical chain with an RMSE improvement of $\times 3.0$ ($1.72\times10^{-4} \rightarrow 5.74\times10^{-5}$), squarely within the held-out range, confirming that the network learns generic soft-tissue anatomy rather than memorising specific geometries (Fig.~\ref{fig:seed9000}).

\begin{figure}[htbp!]
    \centering
    \includegraphics[width=0.95\linewidth]{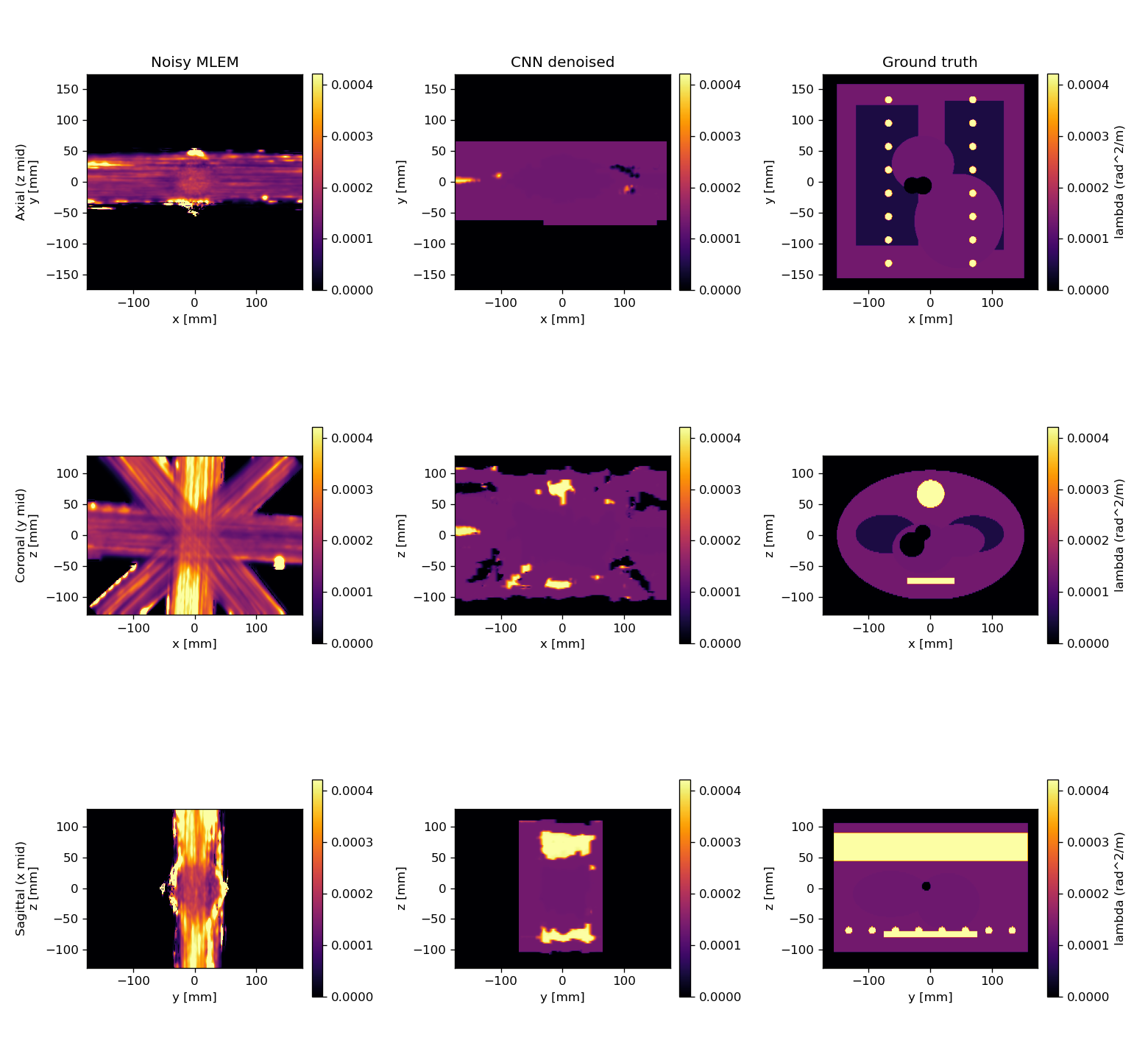}
    \caption{Test phantom seed 45 under real-beam acceptance with $\pm\ang{3}$ angle errors: MLEM input (left), CNN output (centre), ground truth (right). The illumination tube through the heart region and the star-shaped union of the seven beam directions are visible in the MLEM volume; quantitative agreement is evaluated in the common core only.}
    \label{fig:simslices}
\end{figure}

We then repeat the exercise under acceptance-matched illumination, with the simulation restricted to the measured footprint and detector acceptance and the network retrained accordingly --- the configuration that mimics the real scan. Restricting the angular sampling makes the MLEM input both noisier and biased: within the common core the partial-angle sampling inflates the reconstructed $\lambda$ by $+48\%$ relative to the truth. The denoiser removes this bias and recovers the ground-truth core mean to $0.4\%$ ($0.999 \pm 0.004$ across the five test phantoms), while reducing the core RMSE by a factor of $13.6 \pm 4.6$ (Table~\ref{tab:ensemble}, Fig.~\ref{fig:simslices}). The larger RMSE gain than in the full-coverage case simply reflects that the acceptance-restricted MLEM starts much further from the truth; the physically meaningful outcome is that the \emph{absolute} scattering-density scale is recovered correctly in the region that is fully sampled.

\begin{table}[ht!]
\centering
\footnotesize
\begin{tabular}{cccccc}
\toprule
Seed & $N$ muons & MLEM RMSE & CNN RMSE & Gain & CNN/GT core mean \\
\midrule
45 & 339{,}704 & $6.96\times 10^{-5}$ & $6.63\times 10^{-6}$ & $\times 10.5$ & 0.999 \\
46 & 340{,}004 & $5.89\times 10^{-5}$ & $7.51\times 10^{-6}$ & $\times 7.8$  & 1.003 \\
47 & 339{,}934 & $7.50\times 10^{-5}$ & $3.64\times 10^{-6}$ & $\times 20.6$ & 0.992 \\
48 & 339{,}801 & $7.00\times 10^{-5}$ & $4.10\times 10^{-6}$ & $\times 17.1$ & 1.003 \\
49 & 339{,}866 & $6.30\times 10^{-5}$ & $5.37\times 10^{-6}$ & $\times 11.7$ & 0.998 \\
\midrule
mean $\pm$ std & $\sim$340k & $(6.73 \pm 0.57)\times 10^{-5}$ & $(5.45 \pm 1.5)\times 10^{-6}$ & $\times (13.6 \pm 4.6)$ & $0.999 \pm 0.004$ \\
\bottomrule
\end{tabular}
\caption{Production-configuration performance on the five held-out test phantoms: real-beam acceptance, measured-footprint illumination, $\pm\ang{3}$ projection-angle errors, angle-error-trained network. RMSE in \si{\radian\squared\per\metre}, evaluated in the common core.}
\label{tab:ensemble}
\end{table}

\subsection{Soft-tissue reconstruction from the beam data}
\label{sec:realdata}

\begin{figure}[htbp!]
    \centering
    \includegraphics[width=0.9\linewidth]{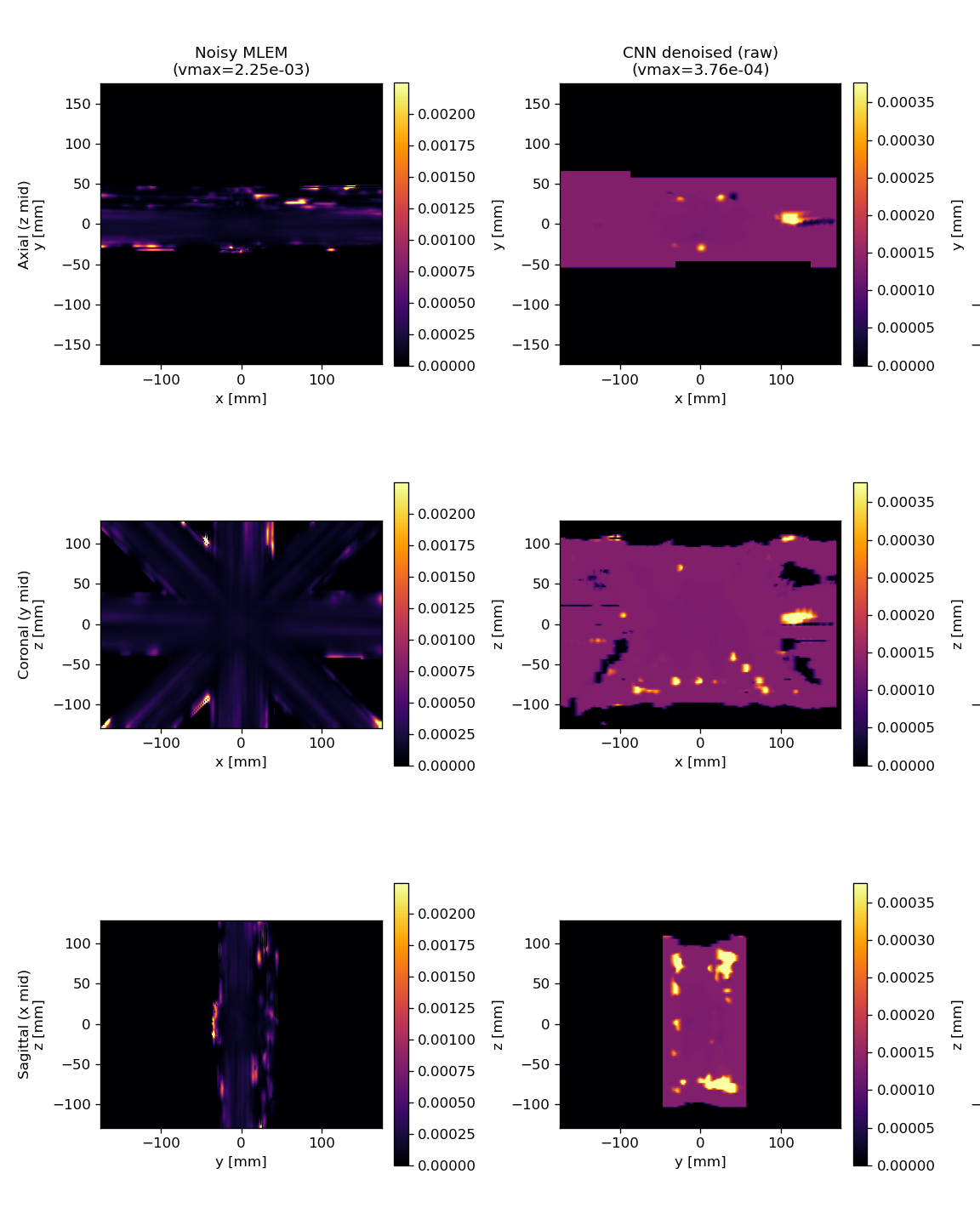}
    \caption{Real-beam reconstruction (default $\theta < \qty{30}{\milli\radian}$ selection): MLEM input (left), raw CNN output (right), in axial, coronal and sagittal views. Within the common core it is quantitatively on the same absolute $\lambda$ scale as the simulation. To be compared with Fig.~\ref{fig:simslices} from simulation.}
    \label{fig:realslices}
\end{figure}

Applying the acceptance-matched, simulation-trained network to the real scan --- with no rescaling or recalibration and the measured MLEM volume fed to the network exactly as reconstructed --- yields a common-core mean scattering density of $1.304\times10^{-4}$~\si{\radian\squared\per\metre}. This agrees at the sub-percent level with both the simulation-ensemble network output ($1.296\times10^{-4}$) and the simulated ground truth ($\approx 1.297\times10^{-4}$; Table~\ref{tab:realresults}), demonstrating that a network trained purely on simulation reproduces the absolute soft-tissue scattering density of real beam data. For the reconstruction from beam data, an angular cut was applied to suppress scattering from hadron contamination or very low energy tail in the beam. Tightening the scattering-angle selection from $\theta<\qty{30}{\milli\radian}$ to $<\qty{20}{\milli\radian}$ shifts the raw MLEM level by $15\%$ but the network output by less than $1\%$, showing that the network and the results are robust. Figure~\ref{fig:realslices} shows the reconstructed volume: within the common core the image is uniform soft tissue, as expected for the homogeneous Alderson heart region, while in the partially sampled region outside the core the ribs, sternum and spine appear at their approximate positions with $\lambda \sim 3.8\times10^{-4}$, close to the simulated cortical-bone value of $\sim 4\times10^{-4}$ --- so even these high-density structures reconstruct near their true scattering density despite the incomplete angular coverage.

\begin{table}[ht!]
\centering
\footnotesize
\begin{tabular}{lcccc}
\toprule
Reconstruction & Muons & MLEM core p50 / p95 & CNN raw core median / mean \\
\midrule
$\theta < \qty{30}{\milli\radian}$ (default) & 293{,}035 & $1.387\times 10^{-4}$ / $5.29\times 10^{-4}$ & $1.294\times 10^{-4}$ / $1.304\times 10^{-4}$ \\
$\theta < \qty{20}{\milli\radian}$ (systematic) & 287{,}681 & $1.184\times 10^{-4}$ / $3.13\times 10^{-4}$ & $1.291\times 10^{-4}$ / $1.290\times 10^{-4}$ \\
\midrule
Simulation reference &&& CNN $1.296\times 10^{-4}$, GT $1.297\times 10^{-4}$ \\
\bottomrule
\end{tabular}
\caption{Real-beam results (Alderson thorax, seven angles, divergence-aligned, raw CNN inference with no rescaling), compared with the simulation ensemble and ground truth in the common core. All values in \si{\radian\squared\per\metre}. p50/p95 and median/mean are percentiles / moments of the core-voxel $\lambda$ distribution.}
\label{tab:realresults}
\end{table}

\FloatBarrier

\section{Discussion}
\label{sec:outlook}

An anthropomorphic thorax phantom was imaged with \qty{2}{\giga\electronvolt\per\clight} muons at the CERN PS T9 beamline at seven projection angles, and reconstructed with an MLEM inversion followed by a 3D U-Net denoiser trained exclusively on simulation. On held-out simulated phantoms the network improves the core RMSE by $13.6 \pm 4.6$ and reproduces the ground-truth scattering density to $0.4\%$; on the fully independent seed-9000 phantom the full-coverage baseline improvement ($\times 3.0$) matches the test ensemble. On real data the raw network output lands within $1\%$ of the simulation-based expectation and is stable to $<1\%$ against the scattering-angle selection.

\subsection{Muon statistics and the density-differentiation floor}
\label{sec:stats}

The number of detected muons per voxel sets the scale of the per-voxel statistical fluctuations and therefore the floor on how finely two soft-tissue densities can be separated. A muon contributes a scattering-angle measurement to every voxel along its longitudinal column, so the relevant quantity is the per-voxel muon count $N$ summed over the seven projection angles. Projected to the phantom plane, the measured beam delivers, per \qtyproduct{1 x 1}{\milli\metre} column and per angle, up to $\approx 76$ muons at the beam-spot maximum and $\approx 44$ averaged over the central $\pm\qty{8}{\milli\metre}$ of the spot, falling to $\approx 11$ over the full illuminated footprint. The fully sampled common core lies in the lower-occupancy part of each footprint, with 6--13 muons per column per angle (mean $\approx 8$). Summed over the seven angles, each core voxel is therefore crossed by $N \approx 40$--$90$ muons (mean $\approx 58$). For comparison, the full-coverage simulation depicted in Section~\ref{sec:performance} generates \num{3e6} muons per angle over the flat \qtyproduct{350 x 350}{\milli\metre} field, i.e.\ $1.225\times10^{5}$ columns of \qtyproduct{1 x 1}{\milli\metre}, giving $\num{3e6}/(1.225\times10^{5}) \approx 25$ muons per column per angle and $N \approx 25\times 7 = 175$ in every voxel; the acceptance-matched simulation, footprint-matched to the measured beam, instead reproduces the real core occupancy ($N \approx 58$).
 
The statistical limit follows directly from the measurement. Each muon yields two independent projected scattering angles $(\Delta x', \Delta y')$, each Gaussian with variance $\sigma^2_\theta \simeq \lambda L$ for a homogeneous path of length $L$; the scattering density is estimated from this variance. The relative uncertainty on a variance estimated from $n$ Gaussian samples is $\sqrt{2/n}$, and with $n = 2N$ angle measurements per voxel this gives
\[
  \frac{\sigma_{\hat\lambda}}{\lambda} \;\simeq\; \sqrt{\frac{2}{2N}} \;=\; \frac{1}{\sqrt{N}}.
\]
A single angle at 25 muons therefore carries a $\approx 20\%$ per-column uncertainty; the full-coverage simulation ($N = 175$) reaches $\approx 8\%$, and the real core and the acceptance-matched simulation ($N \approx 58$) $\approx 13\%$ (10--15\% across the core). These are the information-theoretic floors on a single column's $\lambda$. The MLEM voxel map is noisier still --- its core RMSE is $6.7\times 10^{-5}$, about half the tissue $\lambda$, because the angle-limited inversion amplifies the per-column noise --- and it is this map that the network cleans. The MLEM smoothing ($\sigma = 1$ voxel) and, far more effectively, the learned prior exploit the spatial correlation between neighbouring voxels: on the acceptance-matched test phantoms the production network reduces the core RMSE from $6.73\times 10^{-5}$ ($52\%$ of the tissue $\lambda$) to $5.45\times 10^{-6}$ ($4.2\%$). Averaged over the $\sim$52{,}000-voxel core the mean $\lambda$ is limited only at the $\sim 0.1\%$ level by statistics, so the sub-percent agreement between the real-data mean, the simulation and the ground truth (Section~\ref{sec:realdata}) is a meaningful test of the absolute scale; the $0.4\%$ quoted for the network (see Table~\ref{tab:ensemble}) is the phantom-to-phantom spread of this mean, still well above the statistical floor.
 
The same arithmetic explains why the myocardial defects are not recovered. An air-filled defect of \qtyrange{9}{14}{\milli\litre} is a sphere of \qtyrange{26}{30}{\milli\metre} diameter; a muon crossing it centrally traverses $\approx \qty{27}{\milli\metre}$ of air in place of tissue along a total scattering path of $\approx \qty{250}{\milli\metre}$, so its measured $\theta^2$ is reduced by only $\approx 10\%$. Unlike the homogeneous bulk, where the prior can average over many correlated voxels, a small defect is a localised feature that must be recovered from the raw per-column information, whose floor is $\approx 13\%$ at $N \approx 60$ --- comparable to the $\approx 10\%$ signal. The defect therefore sits at the detection threshold and is largely lost. Increasing the statistics tenfold ($N \approx 600$, per-column floor $\approx 4\%$) would place the defect signal at the edge of detectability.

The radiation exposure associated with muon imaging is intrinsically small. A minimum-ionising GeV muon deposits approximately \SI{2}{\mega\electronvolt\,\gram^{-1}\,\centi\meter^{2}} while traversing tissue. In the fully sampled core of the present measurement, approximately $58$ muons cross each \qtyproduct{1 x 1}{\milli\metre} column summed over the seven projections with a fluence of $\approx\qty{5.8e3}{\per\centi\metre\squared}$, corresponding to a local absorbed dose of about \qty{2}{\micro\gray}. Since the radiation weighting factor for muons is $w_R = 1$, this equals a local equivalent dose of about \qty{2}{\micro\sievert}~\cite{ICRP103}.
Increasing the statistics by an order of magnitude, to the level indicated above for resolving small density variations, scales the dose proportionally, to about \qty{20}{\micro\gray} (\qty{20}{\micro\sievert}). For comparison, a conventional chest radiograph corresponds to an effective dose of approximately \SI{0.1}{\milli\sievert}, while low-dose lung-screening CT and conventional chest CT are typically of order \SI{1.5}{\milli\sievert} and \SI{6}{\milli\sievert}, respectively ~\cite{chestDoseRef}.
These quantities are not directly equivalent, since effective dose includes organ-specific tissue weighting, and a clinical assessment of muon imaging would require a full dosimetric calculation including the spatial fluence distribution and secondary radiation.
Nevertheless, this estimate illustrates the intrinsically low energy deposition associated with GeV-muon imaging.

\subsection{Limited-angle sampling and alignment artefacts}
\label{subsec:align}
The quantitative result holds in the fully sampled, homogeneous common core. The partially sampled peripheral anatomy is a different regime. In the real beam reconstruction the high-density ribs, sternum and spine appear at approximately their correct positions and scattering density but are drawn out into streaks along the beam direction in the sagittal ($y$--$z$) plane (Fig.~\ref{fig:realslices}) that are not present in the reconstruction of the acceptance matched simulation (Fig.~\ref{fig:simslices}). These structures lie outside the seven-fold overlap and are sampled at only one or two projection angles, so the tomographic problem there is limited-angle and poorly constrained. The transverse alignment is also known only at the centimetre scale and the beam drifted between runs, so the seven projections did not all sample the same material and a peripheral structure is caught inconsistently from view to view. Because maximum-likelihood expectation--maximisation conserves the total scattering power, this inconsistent sampling does not corrupt the recovered density scale --- which is why the ribs still reconstruct near the cortical-bone value --- but it does displace and elongate the structures. The simulation shows none of this because this misalignment and drift was not included in the simulation.

This points directly at what a clinical-grade system will require. Diagnostic targets frequently sit in exactly such structured, high-contrast regions --- at organ and tissue boundaries or against bone --- rather than in a homogeneous bulk, so accurate reconstruction there demands both denser angular coverage, to remove the limited-angle ambiguity, and precise geometric alignment. In practice this means a surveyed geometry in which the positions of the trackers and of the patient, and the rotation axis, are known to well below the voxel scale, which is entirely achievable for such devices.

\FloatBarrier

To our knowledge, this is the first validation of soft-tissue anatomical volume imaging with muon scattering tomography at this level of precision. It defines a path for the further optimisation and studies needed to establish MST as a clinical-grade imaging modality. Because the muon scattering density scales essentially with material density, the natural clinical targets for MST are density-contrasted lesions --- a solid tumour or nodule in aerated lung, where the $\lambda$ contrast reaches a factor of a few, or calcified and bony involvement --- whereas near-isodense soft-tissue masses, like the myocardial defects, may be statistics-limited. The trade-off between scanning time and the region of interest can now be studied quantitatively: a small volume imaged with more muons per voxel for higher resolution, against a large volume imaged with fewer muons per voxel for full-body coverage at coarser resolution. Additionally the impact of the beam momentum spread on the achievable imaging resolution shall be included in future campaigns. With the rapid progress of compact laser-driven muon sources~\cite{calvin2023,zhangmuon2025,lblmuons2025}, which promise \unit{\giga\electronvolt} muon fluxes from room-sized facilities, this study establishes low-dose muon tomography of soft tissue as a new and viable application for the emerging laser-accelerated muon landscape.

\section{Methods}
\label{sec:methods}
\subsection{Beamline}
\label{sec:beamline}
 
To assess the feasibility of soft-tissue muon scattering tomography, we performed measurements at the T9 beamline in the East Area of the CERN Proton Synchrotron (PS). T9 delivers secondary beams of hadrons, electrons, and muons with momenta up to \qty{15}{\giga\electronvolt\per\clight}, produced by the \qty{24}{\giga\electronvolt\per\clight} PS proton beam on a production target, with a full momentum acceptance of approximately $\pm 15\%$ around the selected central momentum. For the imaging runs, the beamline was configured for positive muons at \qty{2}{\giga\electronvolt\per\clight}. To generate this beam, an initial \qty{3.6}{\giga\electronvolt\per\clight} mixed hadron beam, comprising protons, pions, and positrons, is transported to a point \qty{25}{\metre} from the production target. Along this path, $\sim$ 12\% of the pions decay to muons carrying \qtyrange{57}{100}{\percent} of the parent momentum. At \qty{25}{\metre}, the hadrons are stopped in a \qty{1}{\metre} iron collimator through which only the muons pass, following which a downstream dipole selects the final muons centered at \qty{2}{\giga\electronvolt\per\clight}. The muon beam produced in this configuration can have a momentum band up to 15\% and an expected hadron contamination of \qtyrange{10}{15}{\percent}. The beam rate was $\sim$ \num{6000} muons per \qty{0.5}{\second} spill over a \qtyproduct{10 x 10}{\centi\metre} scintillator. A dedicated hadron run at \qty{15}{\giga\electronvolt\per\clight} with a momentum spread of $\pm$ 1\% and a flux $\sim$ \num{40000} hadrons per \qty{0.5}{\second} spill over a \qtyproduct{10 x 10}{\centi\metre} scintillator was also recorded for tracker alignment (see Section~\ref{sec:campaign}).

\subsection{Experimental setup}
\label{sec:setup}
 
The tracking setup consists of four Micromegas (MM) trackers with an active area of $8 \times 8$~\unit{\centi\metre\squared} each, read out with trigger-less VMM-based front-end electronics~\cite{Pfeiffer:2021egd}. The MM trackers provide a spatial resolution of the order of $\SI{80}{\micro \meter}$. The detectors are mounted rotated by \ang{45} about the beam axis, so the geometric acceptance projected on the transverse plane is a diamond with $|x| + |y| \leq \qty{56.6}{\milli\metre}$. Two trackers upstream and two downstream of the phantom measure the incoming and outgoing tracks of each muon; the longitudinal positions are
\[
  z_{\mathrm{T1}} = \qty{0.000}{\metre}, \quad
  z_{\mathrm{T2}} = \qty{0.331}{\metre}, \quad
  z_{\mathrm{T3}} = \qty{1.124}{\metre}, \quad
  z_{\mathrm{T4}} = \qty{1.461}{\metre},
\]
with the phantom centre at $z = \qty{0.666}{\metre}$ in the \qty{79.3}{\centi\metre} gap between T2 and T3. These positions were measured to a precision of $\pm\qty{1}{\milli\metre}$. The scattering angle of each muon is obtained from the difference between the upstream (T1--T2) and downstream (T3--T4) track slopes in both transverse projections. The four tracking detectors with the phantom were placed on a horizontal and vertical movable table as shown in Fig.~\ref{fig:phantom_photo} with a positioning accuracy of $\sim$\qty{0.5}{\milli\metre}.

\subsection{Phantom}
\label{sec:phantom}
 
The imaged object is an anthropomorphic Alderson thorax phantom (Radiology Support Devices, RSD), a clinical dosimetry and imaging phantom comprising a tissue-equivalent polyurethane body (density \qty{1.10}{\gram\per\cubic\centi\metre}), asymmetric lungs (\qty{0.40}{\gram\per\cubic\centi\metre}), a heart insert with myocardium- and blood-equivalent compartments, a liver, and a human skeleton (ribs, sternum, spine) in cortical-bone-equivalent material. The phantom was mounted on a rotation platform~\cite{turntable} and imaged at seven projection angles, $\phi \in \{0, 45, 90, 135, 180, 225, 270\}$~degrees, about the vertical (superior--inferior) axis. The rotation was achieved by manually rotating the phantom on the platform with approximately measured angular steps without survey. The achieved precision on the projection angles is therefore expected to be in the range of $\pm$ \qtyrange{1}{3}{\degree}. The beam spot covers only the central mediastinal/heart region of the thorax at each angle (see Section~\ref{sec:imaged}).

\subsection{Data-taking campaign}
\label{sec:campaign}
 
The data were taken in April 2026 over one week. First, on 10~April, with \emph{no phantom installed}, a hadron alignment run and a muon air reference run were recorded. The seven phantom configuration runs followed: \ang{90} on 12~April, \ang{0} on 13~April, and the remaining five angles (\ang{45}, \ang{135}, \ang{180}, \ang{225}, \ang{270}) on 14~April.
 
\begin{figure}[htbp!]
    \centering
    \includegraphics[width=0.6\linewidth]{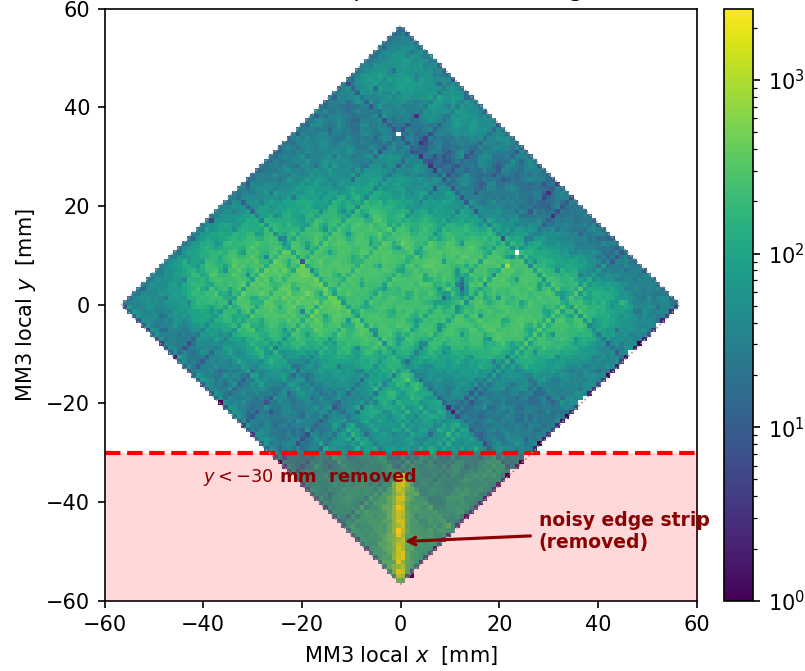}
    \caption{Per-tracker fiducial edge cut, shown for MM3
    (\ang{90} run). Single-hit map: The fiducial cut $y \geq \qty{-30}{\milli\metre}$ (red dashed line) removes the clusters in the shaded region. The analogous cuts on MM1 ($y \geq \qty{-30}{\milli\metre}$),
    MM2 and MM4 ($x \leq \qty{30}{\milli\metre}$) target the equivalent noisy edge of each plane; applied together they retain $93.7\%$ of the four-plane events.}
    \label{fig:fiducial}
\end{figure}

For the event reconstruction, each Micromegas coordinate is read out through 320 strips of \qty{0.25}{\milli\metre} pitch, multiplexed onto 64 VMM channels. Strip signals from a single particle crossing are first grouped in time into per-detector accumulation windows; a window is closed when no further strip fires within \qty{200}{\nano\second}, short enough that pile-up within a window is negligible at the beam rate. Within this window, contiguous fired strips are grouped into clusters, each required to contain at least three strips, which also suppresses ghost combinations from the strip multiplexing. The cluster position is the charge-weighted centroid of its strips and its time the charge-weighted mean of the strip times. Only windows with exactly one $x$- and one $y$-cluster are retained, yielding one unambiguous two-dimensional hit per detector and rejecting pile-up and multi-track ambiguities. Hits in different detectors are matched into events by a timestamp coincidence of $|\Delta t| < \qty{200}{\nano\second}$, a window that comfortably brackets the measured inter-detector coincidence peak (fixed offsets of \qtyrange{-30}{8}{\nano\second} and width $\sigma \approx$ \qtyrange{40}{50}{\nano\second}) while keeping the accidental-coincidence fraction at the percent level at the muon-beam rate; events with hits in all four planes define the track sample used for tomography. Additionally, a per tracker fiducial cut was applied for the clusters to remove noisy strips from the reconstruction as demonstrated in Fig.~\ref{fig:fiducial}. The effect of each cut on the events that survive and finally go into the multiple scattering tomography reconstruction is presented in Table~\ref{tab:selection}. The per-tracker single-hit efficiency and the four-fold selection efficiency are quoted separately because, in the absence of a beam trigger, the individual trackers' hit rates ($10{,}368{,}678$, $5{,}119{,}262$, $4{,}828{,}971$ and $7{,}784{,}232$ windows with at
least one cluster in MM1, MM2, MM3 and MM4 respectively) cannot be combined into an absolute per-detector detection efficiency.

\begin{table}[ht!]
\centering
\footnotesize
\begin{tabular}{lc}
\toprule
Tracker & Single-hit efficiency \\
\midrule
MM1 & $96.3\%$ \\
MM2 & $95.0\%$ \\
MM3 & $93.5\%$ \\
MM4 & $95.3\%$ \\
\bottomrule
\end{tabular}
\hspace{2em}
\begin{tabular}{lcc}
\toprule
Matched-event selection & Removed & Cum. \\
\midrule
Four-fold coincidence & --- & $100\%$ \\
\quad $+$ MM1 fiducial $y\geq\qty{-30}{\milli\metre}$ & $0.36\%$ & $99.6\%$ \\
\quad $+$ MM2 fiducial $x\leq\qty{+30}{\milli\metre}$ & $1.54\%$ & $98.1\%$ \\
\quad $+$ MM3 fiducial $y\geq\qty{-30}{\milli\metre}$ & $3.03\%$ & $95.1\%$ \\
\quad $+$ MM4 fiducial $x\leq\qty{+30}{\milli\metre}$ & $1.53\%$ & $93.7\%$ \\
\bottomrule
\end{tabular}
\caption{Event selection, summed over the seven phantom runs. \emph{Left:} per-tracker
single-hit efficiency (fraction of hit windows with exactly one cluster per plane).
\emph{Right:} the matched-event cut flow relative to the four-fold coincidence sample;
the fiducial cuts jointly retain $93.7\%$.}
\label{tab:selection}
\end{table}

\subsection{Tracker alignment with the hadron beam}
\label{sec:hadronalign}
No precise survey measurements were carried out after installation due to the limited time available for this run. Therefore, the alignment is derived from the beam using the hadron run and assuming straight tracks through the detectors. 
Hadrons at \qty{15}{\giga\electronvolt\per\clight} with \emph{no phantom} in place multiple-scatter far less than \qty{2}{\giga\electronvolt\per\clight} muons and are well focused, so their assumed straight tracks provide a clean geometric reference.

The Gaussian core of each spot defines the per-detector transverse alignment offset in mm: MM1 $(+3.14, +11.64)$, MM2 $(-1.01, +2.80)$, MM3 $(+2.39, +0.76)$, MM4 $(-9.25, +4.34)$~\unit{\milli\metre}. These offsets are subtracted from all hit positions in \emph{all} subsequent runs and constitute the position-alignment correction of the muon analysis.

\subsection{Muon air reference run}
\label{sec:airrun}
 
Following the hadron run, a reference run was recorded with the \emph{muon} beam and \emph{no phantom} installed. This run was used to confirm the effect of the hadron-derived alignment correction on the muon beam, and to measure the instrumental angular response and the angular offset between the tracker arms in the absence of material.
 
Following the hadron position offset corrections the muon beam was found to be still displaced by up to $\approx \qty{10}{\milli\metre}$ from the detector centres, which is expected as the muons and hadrons do not enter the setup along the same trajectory. The hadron beam has a small momentum spread of $\pm 1\%$ and comes from a point-like source at the interaction of the proton beam on the target, whereas the muons, produced in pion decays along the line, have a much larger momentum spread and are steered onto a different trajectory not originating from a point source. A position alignment derived from the hadron beam therefore centres the detectors on the hadron trajectory, and the muons remain off-centre in that frame. For scattering tomography, this residual displacement and tilt are harmless, since only slope \emph{differences} enter the measurement. The offsets from the hadron beam run therefore serve as the reference for aligning the trackers, and a track-slope correction is applied for the scattering measurement as described in Section~\ref{sec:divalign}.

\begin{figure}[htbp!]
    \centering
    \includegraphics[width=\linewidth]{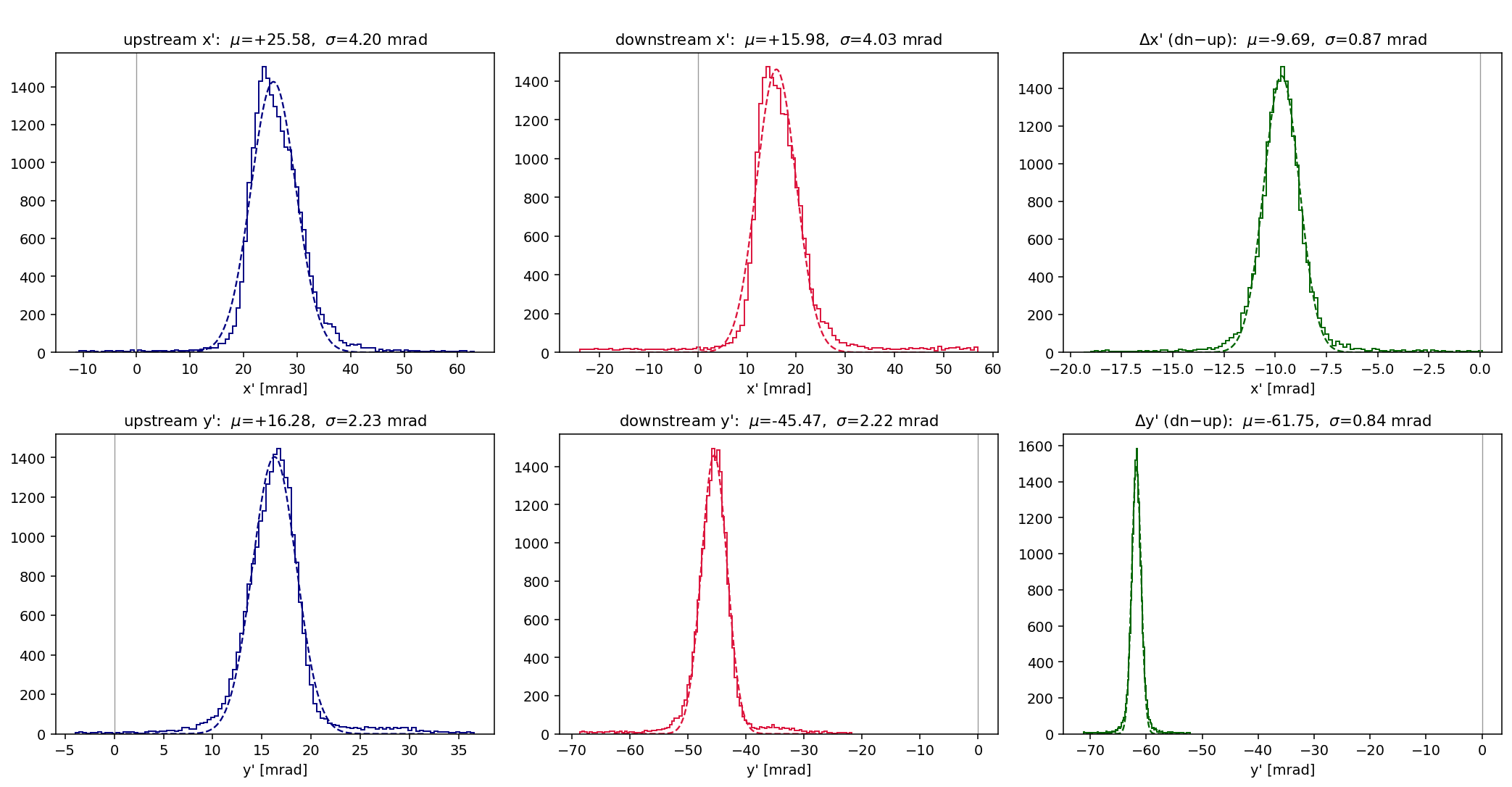}
    \caption{Muon air reference run (10 April 2026, no phantom): upstream and downstream arm slopes and their difference after the hadron position-offset correction, with Gaussian fits. The beam direction differs from the hadron run by $\approx(+23, +6)$~\unit{\milli\radian}, and the arm difference is offset by $(-9.69, -61.75)$~\unit{\milli\radian} --- the angular misalignment corrected by the per-run divergence alignment. The core width $\sigma_{\Delta} \approx \qty{0.85}{\milli\radian}$ is the instrumental angular response for muons.}
    \label{fig:airdiv}
\end{figure}

The width of the muon slope-difference core, $\sigma_{\Delta} \approx \qty{0.85}{\milli\radian}$ per projection as shown in Fig.~\ref{fig:airdiv}, is the in-situ instrumental angular response for muons (including their multiple scattering in the air between the upstream and downstream tracks), to be compared with \qtyrange{4.4}{5.1}{\milli\radian} for the phantom runs, whose width is dominated by genuine multiple scattering in tissue.
 
\subsection{Illuminated area of the phantom}
\label{sec:profiles}
 
\begin{figure}[htbp!]
    \centering
    \includegraphics[width=\linewidth]{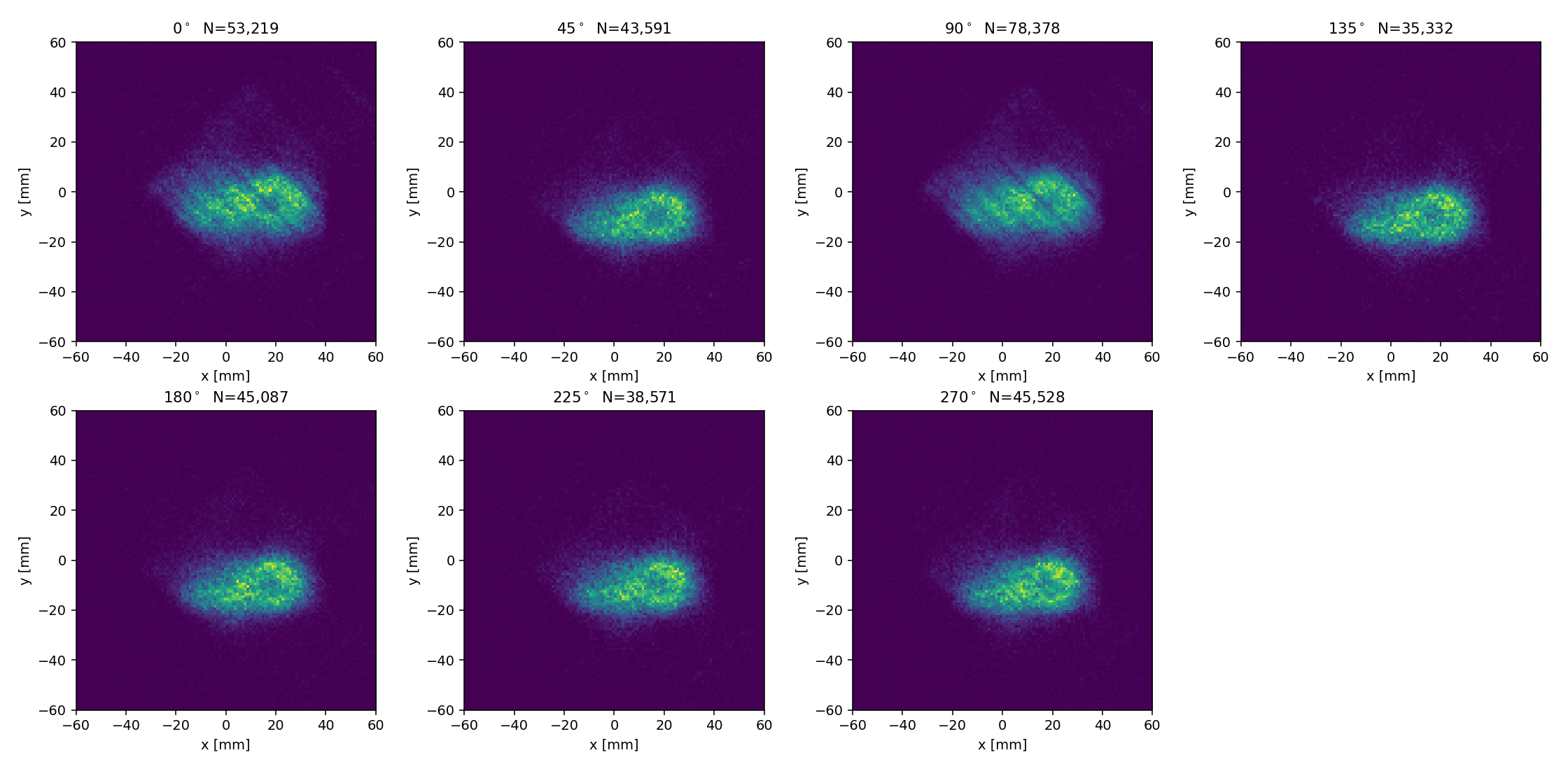}
    \caption{Measured beam footprints at the phantom plane for the seven projection angles, after quality cuts, hadron-run position alignment and divergence alignment. The colour scale is the number of muons per \qtyproduct{1 x 1}{\milli\metre} column.}
    \label{fig:footprints}
\end{figure}

Figure~\ref{fig:footprints} shows the measured muon footprints at the phantom plane ($z = \qty{0.666}{\metre}$), obtained by projecting the upstream track of every accepted event. The beam is not a uniform field but a structured spot of roughly \qtyproduct{60 x 40}{\milli\metre} (full width). The spot differs slightly between the \ang{0}/\ang{90} runs and the other five angles. The beamline configuration was not changed but debugging required the exchange of the readout chips of MM2 detector which caused small movements in the tracker. A vertical movement of the table was also required in between runs to move the setup out of beam due to a parallel data taking campaign. The limited precision of the alignment therefore resulted in run dependent observations. This does not impact the reconstruction, as only the scattering of the muons enter the calculation and a run-wise divergence correction is applied (Section~\ref{sec:divalign}). The footprint --- not the nominal detector acceptance --- defines which voxels of the phantom are actually sampled at each angle; the measured footprints are injected into the training-set construction (see Section~\ref{sec:training}) and constitute training model of the simulation applied to data.
 
\subsection{Divergence alignment of the phantom runs}
\label{sec:divalign}
 
\begin{figure}[htbp!]
    \centering
    \includegraphics[width=\linewidth]{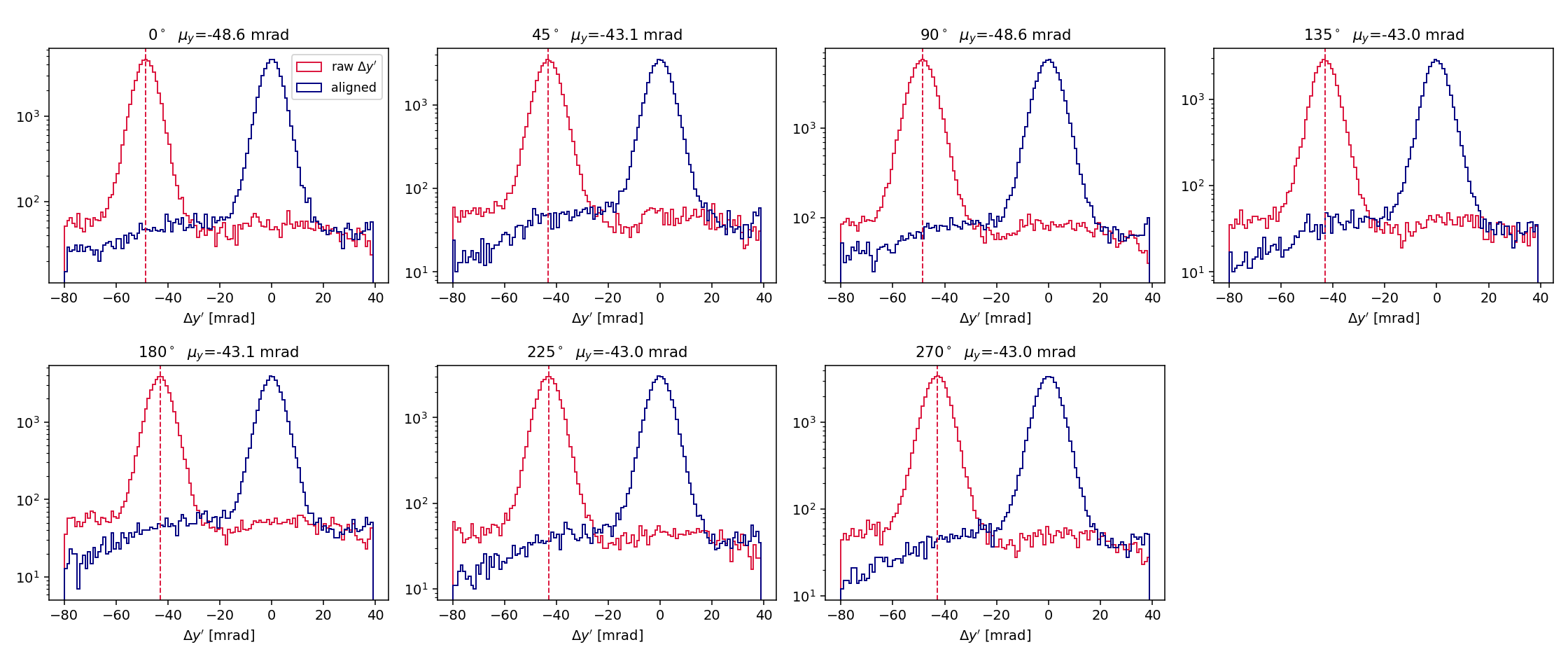}
    \caption{Distribution of the slope difference $\Delta y'$ between the downstream and upstream arms for the seven phantom runs, before (red) and after (blue) the divergence alignment. The dashed line marks the fitted offset $\mu_y$. The offsets fall into two groups (\ang{0}/\ang{90}, taken 12--13 April, vs.\ the five runs of 14 April), consistent with changes to the setup during debugging between the run groups.}
    \label{fig:divalign}
\end{figure}

To confirm consistency between the different projection angle runs, the signed slope-difference distributions $\Delta x' = x'_{\mathrm{down}} - x'_{\mathrm{up}}$ and $\Delta y' = y'_{\mathrm{down}} - y'_{\mathrm{up}}$ of the seven phantom runs were examined: they carry run-dependent constant offsets --- $(-9.7, -48.6)$~\unit{\milli\radian} for the \ang{0}/\ang{90} runs and $(-6.1, -43.0)$~\unit{\milli\radian} for the runs of 14~April (Table~\ref{tab:runs}). A per-run divergence alignment is therefore applied: the robust mean of $(\Delta x', \Delta y')$, clipped at three median absolute deviations (MAD), is computed per configuration and subtracted from the downstream slopes before the scattering angle is formed. Figure~\ref{fig:divalign} shows the $\Delta y'$ distributions before and after the correction. After alignment the scattering-angle median drops from $\sim\qty{50}{\milli\radian}$ to \qtyrange{5.4}{6.3}{\milli\radian}, and a final selection $\theta < \qty{30}{\milli\radian}$ (retaining $86.3\%$ of tracks) removes the residual non-Gaussian tail and any hard-scattering tracks from hadron contamination of the beam as shown in Table~\ref{tab:runs}.

\begin{table}[ht!]
\centering
\footnotesize
\begin{tabular}{llccccc}
\toprule
Run & Acquisition window & $N$ tracks & \multicolumn{2}{c}{Arm angular offset (\si{\milli\radian})} & $\theta$ median & kept by \\
 & (CEST, 2026) & (after cuts) & $\mu_{x'}$ & $\mu_{y'}$ & after align.\ (\si{\milli\radian}) & $\theta<\qty{30}{\milli\radian}$ \\
\midrule
\ang{90}   & 12 Apr, 23:20--04:20\textsuperscript{\dag} & 78{,}378 & $-9.65$ & $-48.61$ & 6.3 & \multirow{7}{*}{86.3\% avg.} \\
\ang{0}    & 13 Apr, 09:14--14:14 & 53{,}219 & $-9.65$ & $-48.55$ & 5.4 & \\
\ang{45}   & 14 Apr, 07:19--12:19 & 43{,}591 & $-6.14$ & $-43.13$ & 5.8 & \\
\ang{135}  & 14 Apr, 12:57--17:57 & 35{,}332 & $-6.18$ & $-43.03$ & 5.7 & \\
\ang{180}  & 14 Apr, 02:02--07:02 & 45{,}087 & $-6.12$ & $-43.09$ & 5.5 & \\
\ang{225}  & 14 Apr, 18:13--23:13 & 38{,}571 & $-6.14$ & $-43.03$ & 5.9 & \\
\ang{270}  & 14 Apr, 23:22--04:22\textsuperscript{\dag} & 45{,}528 & $-6.15$ & $-42.98$ & 6.2 & \\
\bottomrule
\end{tabular}
\caption{The data-taking campaign in chronological order with the measured angular offset between the downstream and upstream tracker arms (Gaussian fit of the slope-difference distributions after position alignment, Sections~\ref{sec:hadronalign}--\ref{sec:divalign}), and the median scattering angle of the phantom runs after divergence alignment. The final $\theta < \qty{30}{\milli\radian}$ selection retains on average $86.3\%$ of the phantom-run tracks ($293{,}035$ muons in total). \textsuperscript{\dag}Acquisition window crosses midnight into the following day.}
\label{tab:runs}
\end{table}
 
\subsection{MLEM scattering-density reconstruction}
\label{sec:mlem}
 
For each muon the upstream and downstream arms are fitted from the two hits each, and the projected scattering angles $(\Delta x', \Delta y')$ are computed. The object volume is discretised on a $350 \times 350 \times 130$ grid of \qtyproduct{1 x 1 x 2}{\milli\metre} voxels ($x$ = left--right, $y$ = superior--inferior = rotation axis, $z$ = anterior--posterior). For each projection angle the voxel grid is rotated about $y$ and the muon paths are traced through the rotated grid, with the expected variance of each muon's scattering angle modelled as $\sigma^2_\theta = \sum_j \lambda_j \ell_{ij}$, where $\ell_{ij}$ is the path length of muon $i$ in voxel $j$. A joint MLEM iteration over all seven angles then updates the voxel scattering densities $\lambda_j$ so as to maximise the likelihood of the observed scattering angles. The iteration count was fixed on the simulation from the convergence of the reconstructed density: the mean (bulk) scattering density rises over the first several MLEM iterations and then plateaus --- changing by less than $1\%$ per iteration beyond ${\sim}15$. Twenty iterations are taken as the point at which the density has converged and the identical count is applied to the real data, followed by a Gaussian smoothing of $\sigma = 1$ voxel. All muons are used with a scattering-angle acceptance of $\theta < \qty{30}{\milli\radian}$ for data (Section~\ref{sec:divalign}) and \qty{150}{\milli\radian} in simulation, where no misalignment or beam-contamination tail exists. An important property of this estimator, used repeatedly below, is that MLEM conserves the total measured scattering power: errors in the assumed projection angles displace $\lambda$ spatially, causing blurring, but cannot change its overall scale. The resulting blur is a tangential smear of order $R\,\delta$ --- about \qtyrange{1}{2}{\milli\metre} at the core radius ($R \approx \qty{40}{\milli\metre}$, for an angle error $\delta$ up to the expected $\pm\ang{3}$ (Section~\ref{sec:angerr})). The reconstruction is implemented in Python, with the sparse system matrix and the joint MLEM update built on NumPy~\cite{numpy2020} and SciPy~\cite{scipy2020} (\texttt{scipy.sparse} for the path-length matrix and \texttt{scipy.ndimage} for the Gaussian smoothing).

\subsection{Simulation campaign}
\label{sec:simulation}
 
The training set is generated with BDSIM~\cite{nevay2020} (Geant4-based~\cite{geant4}), using a parametric model of the Alderson Heart \& Thorax phantom built from the manufacturer's specifications~\cite{rsdtechsheet}: an elliptical-cylinder tissue-equivalent body (\qtyproduct{300 x 210}{\milli\metre} cross-section, \qty{300}{\milli\metre} tall, \qty{1.10}{\gram\per\cubic\centi\metre}), asymmetric lungs (907/\qty{1134}{\milli\litre}, \qty{0.40}{\gram\per\cubic\centi\metre}), an ellipsoidal myocardium shell (\qty{238}{\milli\litre}) with blood-pool chambers (\qty{284}{\milli\litre}), three air-filled myocardial defects of 8.9, 13.5 and \qty{41.7}{\milli\litre}, a liver (\qty{980}{\milli\litre}) and a skeleton (rib tori, spine cylinder, sternum) in cortical bone. The air fill of the myocardial defects represents an optimistic, low-density limiting case. Since the scattering density scales with material density
($\lambda \propto 1/X_0 \propto \rho$), an air cavity is the largest possible density \emph{deficit}. A lesion \emph{denser} than its surroundings carries no such bound: its scattering \emph{excess} $\Delta\lambda = \lambda_{\rm tissue}(\rho/\rho_{\rm tissue}-1)$ grows with density, so a calcified or bony lesion --- or a soft-tissue mass in aerated lung --- produces a larger, more readily detected signal than the air void, whereas a near-isodense soft-tissue tumour remains, like the infarct, statistics-limited. Each of the $50 \times 7$ angle configurations was simulated with the real-beam tracker geometry of Section~\ref{sec:setup}. The simulated beam is \qty{2}{\giga\electronvolt} $\mu^+$ with a \emph{flat} square spatial distribution of $\pm\qty{175}{\milli\metre}$ covering the whole phantom, Gaussian divergence ($\sigma_{x'} = \qty{4}{\milli\radian}$, $\sigma_{y'} = \qty{2.5}{\milli\radian}$ matching the real beam data) and a Gaussian $5\%$ energy spread; \num{3e6} muons are generated per angle so that the muon count per \qtyproduct{1 x 1}{\milli\metre} column ($\approx 25$ per angle) matches the measured occupancy of the real beam in the illuminated region. The flat field is a deliberate super-set of the real illumination: it allows a best-case full-coverage reconstruction to be studied, and the real, narrow illumination is then imposed at reconstruction time by the acceptance filter of Section~\ref{sec:training}. Simulated hits use ideal (true) impact positions; the consequences of neglecting the detector resolution and the beam momentum spectrum are discussed in Section~\ref{sec:systematics}.

\subsection{The learned denoiser}
\label{sec:whydenoiser}
 
Denoising networks are \emph{not} part of the standard MST toolchain. In security and industrial MST the target contrast is enormous: uranium ($\lambda \sim \qty{e-1}{\radian\squared\per\metre}$) against air or organics gives signal-to-background ratios of $10^2$--$10^3$, so even a PoCA point cloud with tens of muons per voxel produces a usable image. In a thorax the situation is inverted. Soft tissues span only $\lambda \approx$ \qtyrange{1.2e-4}{2.1e-4}{\radian\squared\per\metre}, lungs sit at $\approx \qty{0.5e-4}{\radian\squared\per\metre}$, and bone at $\approx \qty{4e-4}{\radian\squared\per\metre}$: the tissue--tissue contrast is a few tens of percent and, on an absolute scale, three to four orders of magnitude below dense-material MST. At the same time, the statistics available per scan are fixed by data from the beam time at $\mathcal{O}(20)$ muons per \qtyproduct{1 x 1}{\milli\metre} column per angle. As quantified in Section~\ref{sec:stats}, the resulting statistical fluctuation of the per-voxel MLEM estimate is comparable to or larger than the anatomical contrast sought: the image information is present but sub-threshold for the human eye and for simple filters. A convolutional-neural-network (CNN) denoiser trained on anatomically realistic phantoms supplies exactly the prior that the likelihood alone cannot: local anatomical plausibility. It is the combination --- an unbiased physics-based likelihood inversion followed by a learned spatial prior --- that makes soft-tissue MST at portable-source statistics feasible. The cost of the learned prior is the risk of hallucination outside its training domain; the acceptance-restricted training of Section~\ref{sec:training} and the transfer validation of Section~\ref{sec:realdata} are designed to control precisely that risk.
 
\subsection{Network and training}
\label{sec:training}
 
The denoiser is a 3D U-Net operating on $16 \times 16 \times N_z$ patches of the MLEM volume, with a logarithmic input normalisation centred on $\lambda_{\mathrm{ref}} = \qty{4e-4}{\radian\squared\per\metre}$. Training pairs are (MLEM reconstruction, analytic ground truth) patches from the fifty simulated phantoms, split \emph{by geometry} into 40 training, 5 validation and 5 test phantoms (seeds 45--49). The network is implemented and trained in PyTorch~\cite{pytorch2019}. The 3D U-Net is a standard, widely validated architecture for volumetric medical-image analysis and restoration~\cite{litjens2017survey,isensee2021nnunet}, and convolutional-network denoising of noisy tomographic reconstructions is itself well established in low-dose X-ray CT~\cite{chen2017reddcnn}.
 
The decisive step for real-data transfer is that the training reconstructions are made under the \emph{measured} real-beam acceptance rather than the idealised flat field:
\begin{enumerate}
  \item \textbf{Detector acceptance cut.} Every simulated muon must traverse the $8\times 8$~\unit{\centi\metre\squared}, \ang{45}-rotated active area (diamond $|x|+|y| \leq \qty{56.6}{\milli\metre}$) at all four tracker planes, exactly as in data.
  \item \textbf{Beam-footprint matching.} The surviving muons are resampled by histogram-ratio accept/reject so that their footprint at the phantom plane matches the \emph{measured} per-angle footprint of Figure~\ref{fig:footprints} (projected to $z = \qty{0.666}{\metre}$ and recentred per angle on the phantom axis, since the phantom, not the beam centroid, defines the imaging frame). Muons are never moved --- only selected --- so the physics of each track is untouched.
  \item \textbf{Core-restricted patches.} Training patches are required to lie at least $50\%$ inside the region imaged at \emph{all} seven angles (the ``common core'', Section~\ref{sec:imaged}), so the network never learns to invent anatomy in unsampled voxels.
\end{enumerate}

\subsubsection{Injected projection-angle errors}
\label{sec:angerr}
 
In the real scan the phantom orientation is known only to a few degrees ($\sim \pm$ \qtyrange{1}{3}{\degree}). Rather than calibrating this away, the network is made robust to it: for every training phantom, each of the seven reconstructions is deliberately performed at a wrong angle $\phi_{\mathrm{recon}} = \phi_{\mathrm{true}} + \delta$, with $\delta$ drawn uniformly in $\pm\ang{3}$ independently per angle (reproducible per-seed random streams), while the ground-truth target remains in the true frame. Two facts make this scheme cheap and effective. First, at these statistics a $\pm\ang{3}$ error degrades the MLEM input only marginally ($\sim 2\%$ in core RMSE): the statistical noise dominates the $\sim\qty{2}{\milli\metre}$ tangential smear that the angle error induces at the core radius. Second, because MLEM conserves the total measured scattering power, angle errors displace $\lambda$ but cannot rescale it, so the target's $\lambda$ scale remains learnable. The angle-error-trained network recovers the \emph{same} core RMSE as the exact-angle network evaluated on exact-angle reconstructions ($6.63\times 10^{-6}$ vs.\ $6.61\times 10^{-6}$ for seed 45): the robustness is acquired at zero performance cost, and the angle-error-trained network is used as the production model throughout.

\subsection{Systematic effects not included in the simulation}
\label{sec:systematics}

Three effects present in the data are absent from, or only approximately captured by, the simulation and hence the network's learned reference: the finite angular resolution of the trackers, the beam momentum spread, and the geometric alignment of the trackers and the phantom. The geometric alignment is discussed in Section~\ref{subsec:align}. The projection-\emph{angle} uncertainty of the manual rotation stage is instead covered by construction (Section~\ref{sec:angerr}).
 
\subsubsection{Detector angular resolution}
\label{sec:sysres}
 
The simulation registers the true muon impact positions, i.e.\ perfect spatial resolution, whereas each Micromegas has a resolution of $\SI{80}{\micro \meter}$.

Each arm slope is formed from two planes separated by $L \approx \qty{0.33}{\metre}$, so the per-arm angular resolution is $\sqrt{2}\,\sigma_{\mathrm{hit}}/L$, and the scattering angle --- the difference of the two arm slopes --- has a resolution $\sigma_{\theta,\mathrm{res}} \approx 2\,\sigma_{\mathrm{hit}}/L \approx \qty{0.48}{\milli\radian}$ per projection. This enters as an additive pedestal on the measured scattering variance,
\[
  \theta_{\mathrm{meas}}^2 \;=\; \theta_{\mathrm{MCS}}^2 \;+\; \sigma_{\theta,\mathrm{res}}^2,
\]
biasing the reconstructed $\lambda$ high. Relative to the phantom scattering widths of \qtyrange{4.4}{5.1}{\milli\radian}, the resolution contributes only $(\sigma_{\theta,\mathrm{res}}/\theta_{\mathrm{MCS}})^2 \approx (0.48/4.7)^2 \approx 1\%$ of the measured variance --- a $\sim 1\%$ $\lambda$ pedestal, well below the $\sim 13\%$ per-voxel statistical floor of the present campaign, and therefore not the limiting effect here.
 
The resolution becomes important once the statistics are increased. As the $1/\sqrt{N}$ floor falls (Section~\ref{sec:outlook}), the resolution pedestal --- not statistics --- sets the smallest resolvable contrast. Since $\sigma_{\theta,\mathrm{res}} \propto \sigma_{\mathrm{hit}}/L$, the pedestal is reduced by better spatial resolution, or by a longer lever arm; this trade-off between angular resolution, acceptance and footprint drives the tracker choice for a higher-statistics campaign.

\subsubsection{Beam momentum spread}
\label{sec:momspread}
 
The momentum spread of the T9 muon beam was not measured during the campaign. The simulation, and hence the network's reference $\lambda$ scale, assumed a Gaussian spread of $\sigma_p/p = 5\%$. Because the scattering density is inferred assuming the nominal momentum and $\lambda \propto \langle p_0^2/p^2 \rangle$, a spread biases the reconstructed scale high by $\langle p_0^2/p^2\rangle - 1 \simeq 3\,(\sigma_p/p)^2$ for a spread of RMS width $\sigma_p/p$. The simulated $5\%$ therefore builds a $+0.8\%$ scale factor into the reference.
 
If the true Gaussian spread were as large as $\sigma_p/p = 15\%$, the same relation gives $\langle p_0^2/p^2\rangle - 1 \approx 3\,(0.15)^2 \approx 7\%$, so the reconstructed $\lambda$ scale would be about $6\%$ higher than the value built into the $5\%$ reference. The network --- having learned the $5\%$ reference --- would pass this through as a corresponding $\approx 6\%$ overestimate of the reconstructed tissue $\lambda$. The observed sub-percent agreement of the real-data core mean with the simulation (Section~\ref{sec:realdata}) already disfavours a spread this large: a $6\%$ excess would be clearly visible in the mean. The data are thus more consistent with an effective spread not much larger than the simulated $5\%$, though a residual few-percent scale uncertainty cannot be excluded without a direct measurement.
 
The effect on the image \emph{resolution}, as opposed to the absolute scale, is negligible: a broader momentum distribution widens the per-muon $\theta^2$ ($\langle p_0^4/p^4\rangle \simeq 1 + 10\,(\sigma_p/p)^2$, a $\sim 20\%$ increase from $5\%$ to $15\%$), inflating the per-voxel statistical error by less than about $10\%$ of itself, invisible next to the $1/\sqrt{N}$ floor. The non-Gaussian low-momentum tail scatters anomalously wide; it is removed by the $\theta < \qty{30}{\milli\radian}$ selection, and the $<1\%$ stability of the network output between the 30 and \qty{20}{\milli\radian} cuts confirms the result does not depend on how aggressively it is cut. A detailed study on the impact of the beam momentum spread on the achievable imaging resolution shall be included in future campaigns. 

\subsection{Estimate of the absorbed dose}

The absorbed dose associated with the muon fluence can be estimated from the mean collisional stopping power of a minimum-ionising GeV muon in tissue, which is approximately~\cite{PDG2024}
\[
\left\langle \frac{dE}{dX} \right\rangle
\simeq
\SI{2}{\mega\electronvolt\,\gram^{-1}\,\centi\meter^{2}}.
\]
For a fluence $\Phi$ of muons per unit area, the energy deposited per unit mass is therefore approximately
\[
D \simeq
\Phi
\left\langle \frac{dE}{dX} \right\rangle .
\]
Using
\[
\SI{1}{\mega\electronvolt\per\gram}
=
1.602\times10^{-10}\ \mathrm{Gy},
\]
this becomes
\[
D\,[\mathrm{Gy}]
\simeq
\Phi\,[\mathrm{cm}^{-2}]
\left\langle \frac{dE}{dX} \right\rangle
[\mathrm{MeV\,g^{-1}\,cm^2}]
\times 1.602\times10^{-10}.
\]

In the fully sampled common core, each $1\times1$~\si{\milli\meter\squared} column is crossed on average by approximately 58 muons over the seven projection angles. This corresponds to a fluence
\[
\Phi \simeq
\frac{58}{0.01\ \mathrm{cm^2}}
=
5.8\times10^{3}\ \mathrm{cm^{-2}},
\]
and hence to an absorbed dose
\[
D
\simeq
5.8\times10^{3}
\times 2
\times 1.602\times10^{-10}
\simeq
1.9\times10^{-6}\ \mathrm{Gy}
=
\SI{1.9}{\micro\gray}.
\]
The same estimate applied to the idealised full-coverage configuration studied in simulation, in which \num{3e6} muons per angle flood the entire
\qtyproduct{350 x 350}{\milli\meter} field, gives $\approx 25$ muons per
$1\times1$~\si{\milli\meter\squared} column per angle and hence $N \approx 175$ over the seven
angles, a fluence
\[
\Phi \simeq \frac{175}{0.01\ \mathrm{cm^2}} = 1.75\times10^{4}\ \mathrm{cm^{-2}},
\]
and an absorbed dose
\[
D \simeq 1.75\times10^{4}\times 2 \times 1.602\times10^{-10}
\simeq 5.6\times10^{-6}\ \mathrm{Gy} = \SI{5.6}{\micro\gray}.
\]
In this case, the dose is delivered approximately uniformly across the whole phantom
cross-section.

An order-of-magnitude increase in statistics would therefore correspond to an absorbed dose of approximately \SI{20}{\micro\gray}.
For muons, the radiation weighting factor is $w_R=1$~\cite{ICRP103}, so these values correspond numerically to local equivalent doses of approximately \SI{2}{\micro\sievert} and \SI{20}{\micro\sievert}, respectively.

This estimate neglects the spatial non-uniformity of the beam, secondary particle production, and organ-specific tissue weighting, and should therefore be interpreted as an order-of-magnitude estimate of the local dose rather than as a clinical effective-dose calculation.

\section*{Acknowledgements}
We thank the CERN PS and East Area operations teams for the delivery and support of the T9 beam
during the campaign, and DRD1, especially David Jose Gaspar Marques and Karl Fl\"{o}thner, for their support with the
VMM readout. We thank Dr. Habib Zaidi of H\^opitaux Universitaires de Genève (HUG) for kindly lending us the phantom, and Valentina Garimbotto, Irene Bürgler and Lars Gerchow for putting us in contact with him. P.C and B.B.O gratefully acknowledge the support of the Swiss National Science Foundation (SNSF) under Grants No. 216602 and No. 232705.

\bibliographystyle{unsrt}
\bibliography{references}

@article{borozdin2003,
  author  = {K.~N. Borozdin and G.~E. Hogan and C. Morris and W.~C. Priedhorsky and A. Saunders and L.~J. Schultz and M.~E. Teasdale},
  title   = {Radiographic imaging with cosmic-ray muons},
  journal = {Nature},
  volume  = {422},
  pages   = {277},
  year    = {2003}
}

@article{procureur2018,
  author  = {S. Procureur},
  title   = {Muon imaging: Principles, technologies and applications},
  journal = {Nucl. Instrum. Meth. A},
  volume  = {878},
  pages   = {169--179},
  year    = {2018}
}

@article{bonechi2020,
  author  = {L. Bonechi and R. D'Alessandro and A. Giammanco},
  title   = {Atmospheric muons as an imaging tool},
  journal = {Reviews in Physics},
  volume  = {5},
  pages   = {100038},
  year    = {2020}
}

@article{morishima2017,
  author  = {K. Morishima and others},
  title   = {Discovery of a big void in Khufu's Pyramid by observation of cosmic-ray muons},
  journal = {Nature},
  volume  = {552},
  pages   = {386--390},
  year    = {2017}
}

@article{schultz2007,
  author  = {L.~J. Schultz and G.~S. Blanpied and K.~N. Borozdin and A.~M. Fraser and N.~W. Hengartner and A.~V. Klimenko and C.~L. Morris and C. Orum and M.~J. Sossong},
  title   = {Statistical Reconstruction for Cosmic Ray Muon Tomography},
  journal = {IEEE Trans. Image Process.},
  volume  = {16},
  number  = {8},
  pages   = {1985--1993},
  year    = {2007}
}

@article{morrisphantom2025,
  author  = {C.~L. Morris and J.~O. Perry and others},
  title   = {Cosmic ray radiography of a human phantom},
  journal = {J. Appl. Phys.},
  volume  = {137},
  pages   = {124503},
  year    = {2025},
  note    = {arXiv:2106.01542}
}

@article{lblmuons2025,
  author  = {D. Terzani and others},
  title   = {Measurement of directional muon beams generated at the Berkeley Lab Laser Accelerator},
  journal = {Phys. Rev. Accel. Beams},
  year    = {2025},
  note    = {arXiv:2411.02321.}
}

@article{giomataris1996,
  author  = {Y. Giomataris and Ph. Rebourgeard and J.~P. Robert and G. Charpak},
  title   = {MICROMEGAS: a high-granularity position-sensitive gaseous detector for high particle-flux environments},
  journal = {Nucl. Instrum. Meth. A},
  volume  = {376},
  pages   = {29--35},
  year    = {1996}
}

@article{Banerjee:2017mdu,
    author = "Banerjee, D. and others",
    title = "{Performance of Multiplexed XY Resistive Micromegas detectors in a high intensity beam}",
    eprint = "1708.04087",
    archivePrefix = "arXiv",
    primaryClass = "physics.ins-det",
    doi = "10.1016/j.nima.2017.10.067",
    journal = "Nucl. Instrum. Meth. A",
    volume = "881",
    pages = "72--81",
    year = "2018"
}

@article{Pfeiffer:2021egd,
    author = "Pfeiffer, D. and others",
    title = "{Rate-capability of the VMM3a front-end in the RD51 Scalable Readout System}",
    eprint = "2109.10287",
    archivePrefix = "arXiv",
    primaryClass = "physics.ins-det",
    doi = "10.1016/j.nima.2022.166548",
    journal = "Nucl. Instrum. Meth. A",
    volume = "1031",
    pages = "166548",
    year = "2022"
}

@article{nevay2020,
  author  = {L.~J. Nevay and others},
  title   = {BDSIM: An accelerator tracking code with particle--matter interactions},
  journal = {Comput. Phys. Commun.},
  volume  = {252},
  pages   = {107200},
  year    = {2020}
}

@article{geant4,
  author  = {S. Agostinelli and others},
  title   = {Geant4---a simulation toolkit},
  journal = {Nucl. Instrum. Meth. A},
  volume  = {506},
  pages   = {250--303},
  year    = {2003}
}

@inproceedings{bernhard2019eastarea,
  author    = {J. Bernhard and others},
  title     = {The Beamlines of the CERN East Area Renovation Project},
  booktitle = {Proc. IPAC},
  year      = {2019},
  note      = {JACoW, CERN-ACC-2019; CDS record 2664974}
}

@article{highland1975,
  author  = {V.~L. Highland},
  title   = {Some practical remarks on multiple scattering},
  journal = {Nucl. Instrum. Meth.},
  volume  = {129},
  pages   = {497--499},
  year    = {1975}
}

@article{lynch1991,
  author  = {G.~R. Lynch and O.~I. Dahl},
  title   = {Approximations to multiple Coulomb scattering},
  journal = {Nucl. Instrum. Meth. B},
  volume  = {58},
  pages   = {6--10},
  year    = {1991}
}

@article{sheppvardi1982,
  author  = {L.~A. Shepp and Y. Vardi},
  title   = {Maximum Likelihood Reconstruction for Emission Tomography},
  journal = {IEEE Trans. Med. Imaging},
  volume  = {1},
  pages   = {113--122},
  year    = {1982}
}

@inproceedings{ronneberger2015,
  author    = {O. Ronneberger and P. Fischer and T. Brox},
  title     = {U-Net: Convolutional Networks for Biomedical Image Segmentation},
  booktitle = {MICCAI 2015, LNCS 9351},
  pages     = {234--241},
  year      = {2015}
}

@article{xu2025hiafmuon,
  author        = {Yu Xu and others},
  title         = {The Feasibility Study of the {GeV}-Energy Muon Source Based on {HIAF}},
  journal       = {arXiv preprint arXiv:2502.20915},
  year          = {2025},
  eprint        = {2502.20915},
  archivePrefix = {arXiv},
  primaryClass  = {physics.acc-ph}
}

@article{machida2012emma,
  author  = {S. Machida and others},
  title   = {Acceleration in the linear non-scaling fixed-field alternating-gradient accelerator {EMMA}},
  journal = {Nature Physics},
  volume  = {8},
  pages   = {243--247},
  year    = {2012},
  doi     = {10.1038/nphys2179}
}

@article{knecht2021himb,
  author        = {A. Knecht and others},
  title         = {Science Case for the new High-Intensity Muon Beams {HIMB} at {PSI}},
  journal       = {arXiv preprint arXiv:2111.05788},
  year          = {2021},
  eprint        = {2111.05788},
  archivePrefix = {arXiv},
  primaryClass  = {hep-ex}
}

@inproceedings{cicek2016,
  author    = {{\"O}. {\c{C}}i{\c{c}}ek and A. Abdulkadir and S.~S. Lienkamp and T. Brox and O. Ronneberger},
  title     = {3D U-Net: Learning Dense Volumetric Segmentation from Sparse Annotation},
  booktitle = {MICCAI 2016, LNCS 9901},
  pages     = {424--432},
  year      = {2016}
}

@misc{rsdtechsheet,
  author       = {{Radiology Support Devices Inc.}},
  title        = {Alderson Heart \& Thorax Phantom --- Technical Sheet},
  howpublished = {\url{https://rsdphantoms.com/}},
  year         = {2024}
}

@misc{turntable,
  author       = {{JEDFORE}},
  title        = {360 Degree Turntable with Steel Ball Bearings, Heavy Duty Multi-Function for Large TV, Computer Monitor, 12 Inch},
  howpublished = {\url{https://www.amazon.fr/dp/B01N2HKIV9?ref=ppx_yo2ov_dt_b_fed_asin_title&th=1}},
  year         = {2026}
}

@article{zhangmuon2025,
  author  = {F. Zhang and L. Deng and Y. Ge and others},
  title   = {Proof-of-principle demonstration of muon production with an ultrashort high-intensity laser},
  journal = {Nature Physics},
  volume  = {21},
  pages   = {1050--1056},
  year    = {2025}
}

@article{calvin2023,
  author  = {L. Calvin and P. Tomassini and D. Doria and D. Martello and R.~M. Deas and G. Sarri},
  title   = {Laser-driven muon production for material inspection and imaging},
  journal = {Frontiers in Physics},
  volume  = {11},
  pages   = {1177486},
  year    = {2023}
}

@article{numpy2020,
  author  = {C. R. Harris and K. J. Millman and S. J. van der Walt and others},
  title   = {Array programming with {NumPy}},
  journal = {Nature},
  volume  = {585},
  pages   = {357--362},
  year    = {2020},
  doi     = {10.1038/s41586-020-2649-2}
}

@article{scipy2020,
  author  = {P. Virtanen and R. Gommers and T. E. Oliphant and others},
  title   = {{SciPy} 1.0: fundamental algorithms for scientific computing in {Python}},
  journal = {Nature Methods},
  volume  = {17},
  pages   = {261--272},
  year    = {2020},
  doi     = {10.1038/s41592-019-0686-2}
}

@inproceedings{pytorch2019,
  author    = {A. Paszke and S. Gross and F. Massa and others},
  title     = {{PyTorch}: An Imperative Style, High-Performance Deep Learning Library},
  booktitle = {Advances in Neural Information Processing Systems 32 (NeurIPS)},
  pages     = {8024--8035},
  year      = {2019}
}

@article{litjens2017survey,
  author  = {G. Litjens and T. Kooi and B. E. Bejnordi and others},
  title   = {A survey on deep learning in medical image analysis},
  journal = {Medical Image Analysis},
  volume  = {42},
  pages   = {60--88},
  year    = {2017},
  doi     = {10.1016/j.media.2017.07.005}
}

@article{isensee2021nnunet,
  author  = {F. Isensee and P. F. Jaeger and S. A. A. Kohl and J. Petersen and K. H. Maier-Hein},
  title   = {{nnU-Net}: a self-configuring method for deep learning-based biomedical image segmentation},
  journal = {Nature Methods},
  volume  = {18},
  pages   = {203--211},
  year    = {2021},
  doi     = {10.1038/s41592-020-01008-z}
}

@article{chen2017reddcnn,
  author  = {H. Chen and Y. Zhang and M. K. Kalra and others},
  title   = {Low-Dose {CT} with a Residual Encoder-Decoder Convolutional Neural Network},
  journal = {IEEE Transactions on Medical Imaging},
  volume  = {36},
  number  = {12},
  pages   = {2524--2535},
  year    = {2017},
  doi     = {10.1109/TMI.2017.2715284}
}

@article{yukawa1935,
  author  = {Yukawa, Hideki},
  title   = {On the Interaction of Elementary Particles. I},
  journal = {Proc. Phys.-Math. Soc. Japan},
  volume  = {17},
  pages   = {48--57},
  year    = {1935},
  doi     = {10.11429/ppmsj1919.17.0_48}
}

@article{neddermeyer1937,
  author  = {Neddermeyer, S. H. and Anderson, C. D.},
  title   = {Note on the Nature of Cosmic-Ray Particles},
  journal = {Phys. Rev.},
  volume  = {51},
  pages   = {884--886},
  year    = {1937},
  doi     = {10.1103/PhysRev.51.884}
}

@article{street1937,
  author  = {Street, J. C. and Stevenson, E. C.},
  title   = {New Evidence for the Existence of a Particle of Mass Intermediate Between the Proton and Electron},
  journal = {Phys. Rev.},
  volume  = {52},
  pages   = {1003--1004},
  year    = {1937},
  doi     = {10.1103/PhysRev.52.1003}
}

@article{conversi1947,
  author  = {Conversi, M. and Pancini, E. and Piccioni, O.},
  title   = {On the Disintegration of Negative Mesons},
  journal = {Phys. Rev.},
  volume  = {71},
  pages   = {209--210},
  year    = {1947},
  doi     = {10.1103/PhysRev.71.209}
}

@article{lattes1947,
  author  = {Lattes, C. M. G. and Occhialini, G. P. S. and Powell, C. F.},
  title   = {Observations on the Tracks of Slow Mesons in Photographic Emulsions},
  journal = {Nature},
  volume  = {160},
  pages   = {453--456},
  year    = {1947},
  doi     = {10.1038/160453a0}
}

@article{janka2022muonium,
  author  = {Janka, Gianluca and Ohayon, Ben and Cortinovis, Irene and
             Burkley, Zak and de Sousa Borges, Lucas and Depero, Emilio and
             Golovizin, Artem and Ni, Xiaojie and Salman, Zaher and
             Suter, Andreas and Prokscha, Thomas and Crivelli, Paolo},
  title   = {Measurement of the transition frequency from
             $2S_{1/2}, F=0$ to $2P_{1/2}, F=1$ states in Muonium},
  journal = {Nature Communications},
  volume  = {13},
  pages   = {7273},
  year    = {2022},
  doi     = {10.1038/s41467-022-34672-0}
}

@article{calibbi2018clfv,
  author  = {Calibbi, Lorenzo and Signorelli, Giovanni},
  title   = {Charged Lepton Flavour Violation: An Experimental and Theoretical Introduction},
  journal = {Riv. Nuovo Cim.},
  volume  = {41},
  number  = {2},
  pages   = {71--174},
  year    = {2018},
  doi     = {10.1393/ncr/i2018-10144-0},
  eprint  = {1709.00294},
  archivePrefix = {arXiv},
  primaryClass  = {hep-ph}
}

@article{muong2final,
  author  = {{Muon g-2 Collaboration}},
  title   = {Measurement of the Positive Muon Anomalous Magnetic Moment to 127 ppb},
  journal = {Phys. Rev. Lett.},
  volume  = {135},
  pages   = {101802},
  year    = {2025},
  doi     = {10.1103/7clf-sm2v}
}

@article{webber2011mulan,
  author  = {Webber, D. M. and Tishchenko, V. and Peng, Q. and Battu, S. and
             Carey, R. M. and Chitwood, D. B. and Crnkovic, J. and
             Debevec, P. T. and Dhamija, S. and Earle, W. and others},
  collaboration = {MuLan Collaboration},
  title   = {Measurement of the Positive Muon Lifetime and Determination of the Fermi Constant to Part-per-Million Precision},
  journal = {Phys. Rev. Lett.},
  volume  = {106},
  pages   = {041803},
  year    = {2011},
  doi     = {10.1103/PhysRevLett.106.041803},
  eprint  = {1010.0991},
  archivePrefix = {arXiv},
  primaryClass  = {hep-ex}
}

@article{PDG2024,
  author        = {Navas, S. and others},
  collaboration = {Particle Data Group},
  title         = {Review of Particle Physics},
  journal       = {Phys. Rev. D},
  volume        = {110},
  number        = {3},
  pages         = {030001},
  year          = {2024},
  doi           = {10.1103/PhysRevD.110.030001}
}

@article{ICRP103,
  author  = {{International Commission on Radiological Protection}},
  title   = {The 2007 Recommendations of the International Commission on Radiological Protection},
  journal = {Ann. ICRP},
  volume  = {37},
  number  = {2--4},
  year    = {2007},
  note    = {ICRP Publication 103}
}

@article{hillier2022musr,
  author  = {Hillier, A. D. and others},
  title   = {Muon spin spectroscopy},
  journal = {Nature Reviews Methods Primers},
  volume  = {2},
  pages   = {4},
  year    = {2022},
  doi     = {10.1038/s43586-021-00089-0}
}

@article{andreev2024na64mu,
  author        = {Andreev, Yu. M. and others},
  collaboration = {NA64 Collaboration},
  title         = {First Results in the Search for Dark Sectors at NA64 with the CERN SPS High Energy Muon Beam},
  journal       = {Phys. Rev. Lett.},
  volume        = {132},
  number        = {21},
  pages         = {211803},
  year          = {2024},
  doi           = {10.1103/PhysRevLett.132.211803},
  eprint        = {2401.01708},
  archivePrefix = {arXiv},
  primaryClass  = {hep-ex}
}

@article{schuhmann2025helion,
  author  = {Schuhmann, Karsten and Fernandes, Luis M. P. and Nez, Fran{\c{c}}ois
             and Abdou Ahmed, Marwan and Amaro, Fernando D. and Amaro, Pedro
             and Biraben, Fran{\c{c}}ois and Chen, Tzu-Ling and Covita, Daniel S.
             and Dax, Andreas J. and others},
  title   = {The helion charge radius from laser spectroscopy of muonic helium-3 ions},
  journal = {Science},
  volume  = {388},
  number  = {6749},
  pages   = {854--858},
  year    = {2025},
  doi     = {10.1126/science.adj2610}
}

@article{strasser2025museum,
  author  = {Strasser, Patrick and Abe, Mitsushi and Asai, Kanta and
             Fukumura, Seiso and Fushihara, Mahiro and Goto, Yu and
             Ino, Takashi and Iwai, Ryoto and Kanda, Sohtaro and
             Kawamura, Shiori and Kitaguchi, Masaaki and Nishimura, Shoichiro
             and Oku, Takayuki and Okudaira, Takuya and Powell, Adam and
             Sasaki, Ken-ichi and Shimizu, Hirohiko M. and
             Shimomura, Koichiro and Tada, Hiroki and Torii, Hiroyuki A.
             and Yamanaka, Takashi and Yamazaki, Takayuki},
  title   = {Precision measurements of muonium and muonic helium hyperfine structure at J-PARC},
  journal = {Eur. Phys. J. D},
  volume  = {79},
  pages   = {20},
  year    = {2025},
  doi     = {10.1140/epjd/s10053-025-00959-2}
}

@article{chestDoseRef,
  author  = {Mettler, Fred A. and Huda, Walter and Yoshizumi, Terry T. and Mahesh, Mahadevappa},
  title   = {Effective Doses in Radiology and Diagnostic Nuclear Medicine: A Catalog},
  journal = {Radiology},
  volume  = {248},
  number  = {1},
  pages   = {254--263},
  year    = {2008},
  doi     = {10.1148/radiol.2481071451}
}
 
\end{document}